\documentclass[aps,prd,twocolumn,showpacs,floatfix,preprintnumbers,amsfont,amsmath,amssymb,nofootinbib, superscriptaddress]{revtex4}

\usepackage{color}
\usepackage[font=small]{caption,subfig}
\usepackage{hyperref}
\usepackage[normalem]{ulem}
\usepackage{lipsum}
\usepackage{url}
\usepackage{float}
\usepackage{ctable}
\usepackage{graphicx}
\usepackage[font=small,
   justification=justified,
   format=plain]{caption} 

\newcommand{\refeq}[1]{Eq.~(\ref{eq:#1})}          
          
\newcommand{\reffig}[1]{Fig.~\ref{fig:#1}}          
\newcommand{\refsec}[1]{Sec.~\ref{sec:#1}}

\def\VEV#1{\langle #1 \rangle}
\def\VEVZ#1{\left( #1 \right)}

\def\bfy{\boldsymbol{y}}

\def\bfJ{\boldsymbol{J}}
\def\bfx{\boldsymbol{x}}
\def\bfv{\boldsymbol{v}}
\def\bfu{\boldsymbol{u}}

\def\calP{\mathcal{P}}
\def\calV{\mathcal{V}}
\def\calS{\mathcal{S}}

\newcommand{\be}{\begin{equation}}
\newcommand{\ee}{\end{equation}}
\newcommand{\ba}{\begin{eqnarray}}
\newcommand{\ea}{\end{eqnarray}}
\newcommand{\en}{\nonumber\\}

\newcommand{\mnras}{Mon. Not. R. Astron. Soc.}

\DeclareMathOperator*{\argmax}{arg\,max}

\allowdisplaybreaks 

\begin{document}

\title{Detecting Lensing-Induced Diffraction in Astrophysical Gravitational Waves}

\author{Liang Dai}
\thanks{NASA Einstein Fellow}
\email{ldai@ias.edu}
\affiliation{\mbox{School of Natural Sciences, Institute for Advanced Study, Princeton, New Jersey 08540, USA}}

\author{Shun-Sheng Li}
\affiliation{\mbox{National Astronomical Observatories, Chinese Academy of Sciences, Beijing 100101, China}}
\affiliation{\mbox{School of Astronomy and Space Science, University of Chinese Academy of Sciences, Beijing 100049, China}}

\author{Barak Zackay}
\affiliation{\mbox{School of Natural Sciences, Institute for Advanced Study, Princeton, New Jersey 08540, USA}}

\author{Shude Mao}
\affiliation{\mbox{Physics Department and Tsinghua Centre for Astrophysics, Tsinghua University, Beijing 100084, China}}
\affiliation{\mbox{National Astronomical Observatories, Chinese Academy of Sciences, Beijing 100101, China}}
\affiliation{\mbox{Jodrell Bank Centre for Astrophysics, School of Physics and Astronomy, University of Manchester, Manchester M13 9PL, UK}}

\author{Youjun Lu}
\affiliation{\mbox{National Astronomical Observatories, Chinese Academy of Sciences, Beijing 100101, China}}
\affiliation{\mbox{School of Astronomy and Space Science, University of Chinese Academy of Sciences, Beijing 100049, China}}

\date{\today}


\begin{abstract}

Gravitational waves emitted from compact binary coalescence can be subject to wave diffraction if they are gravitationally lensed by an intervening mass clump whose Schwarzschild timescale matches the wave period. Waves in the ground-based frequency band $f\sim 10$--$10^3\,$Hz are sensitive to clumps with masses $M_E \sim 10^2$--$10^3\,M_\odot$ enclosed within the impact parameter. These can be the central parts of low mass $M_L \sim 10^3$--$10^6\,M_\odot$ dark matter halos, which are predicted in Cold Dark Matter scenarios but are challenging to observe. Neglecting finely-tuned impact parameters, we focus on lenses aligned generally on the Einstein scale for which multiple lensed images may not form in the case of an extended lens. In this case, diffraction induces amplitude and phase modulations whose sizes $\sim 10\%$--$20\%$ are small enough so that standard matched filtering with unlensed waveforms do not degrade, but are still detectable for events with high signal-to-noise ratio. We develop and test an agnostic detection method based on dynamic programming, which does not require a detailed model of the lensed waveforms. For pseudo-Jaffe lenses aligned up to the Einstein radius, we demonstrate that a pair of fully upgraded aLIGO/Virgo detectors can extract diffraction imprints from binary black hole mergers out to $z_s \sim 0.2$--$0.3$. The prospect will improve dramatically for a third-generation detector for which binary black hole mergers out to $z_s \sim 2$--$4$ will all become valuable sources.

\end{abstract}


\maketitle


\section{Introduction}
\label{sec:intro}

Recent detection of gravitational wave (GW) signatures from compact binary coalescence with the ground-based observatory network aLIGO/Virgo has opened up a new window into the Universe~\citep{Abbott:2016blz, Abbott:2016nmj, Abbott:2017vtc, TheLIGOScientific:2017qsa, Abbott:2017gyy, abbott2017gw170817}. Large number of events from an increased volume are expected after aLIGO/Virgo undergo major upgrade and after KAGRA~\cite{Aso:2013eba} and LIGO-India~\cite{Unnikrishnan:2013qwa} join the network in the near future.

GWs can be gravitationally lensed if the line of sight is perturbed by a mass clump such as the dark matter (DM) halo associated with a galaxy or a galaxy cluster~\citep{Wang:1996as, Piorkowska:2013eww, Biesiada:2014kwa, Dai:2016igl, Ng:2017yiu, Smith:2017mqu, Li:2018prc, Broadhurst:2018saj, Oguri:2018muv}. At cosmological distances $z \simeq 1$, about $10^{-3}$ of the events would be strongly lensed by intervening galaxies. If observed, these special events can be used to probe cosmology~\citep{sereno2011cosmography, Liao:2017ioi, doi:10.1093/mnras/stx2210} or to constrain fundamental physics~\citep{PhysRevLett.118.091101, PhysRevLett.118.091102}.

In contrast to the galactic mass scale $M_L \gtrsim 10^{10}\,M_\odot$, the lumpiness of the Universe on smaller mass scales are empirically less understood. In the Cold Dark Matter (CDM) paradigm, DM halos are predicted to span a mass range across many orders of magnitude $M_L \sim 10^{-6}$--$10^{15}\,M_\odot$~\citep{1981ApJ...250..423D, blumenthal1982galaxy, blumenthal1984formation, 1985ApJ...292..371D}. In alternative scenarios, formation of low mass clumps may be suppressed or prohibited, such as in the case of Warm Dark Matter~\citep{PhysRevD.71.063534, colin2000substructure, bode2001halo}, or Bosonic Dark Matter with a macroscopic de Broglie wavelength~\citep{PhysRevD.28.1243, PhysRevD.50.3650, PhysRevLett.85.1158, goodman2000repulsive, peebles2000fluid, PhysRevD.95.043541}. For testing those models, strong lensing of distant electromagnetic sources have been considered as powerful tools to probe halos of low mass scales $M_L \sim 10^6$--$10^9\,M_\odot$, mainly residing in intervening galactic halos~\citep{Mao:1997ek, metcalf2001compound, dalal2002direct, xu2009effects,  vegetti2010detection, xu2010substructure, xu2012effects, vegetti2012gravitational, hezaveh2013dark, xu2015well, Cyr-Racine:2015jwa, hezaveh2016detection, 2017MNRAS.471.2224N, 2017JCAP...05..037B, asadi2017probing} or cluster halos~\citep{Diego:2017drh, Venumadhav:2017pps, Oguri:2017ock, Dai:2018mxx} as substructure.

When the Schwarzschild time corresponding to the lens mass is comparable to the wave period, wave diffraction effects become important~\cite{Takahashi:2003ix, Takahashi:2005ug, Takahashi:2016jom}. In this paper, we focus on the frequency band of ground-based detectors $f \sim 10$--$10^3\,$Hz, which points toward an intriguing mass scale $M_E \sim 10^2$--$10^3\,M_\odot$ enclosed within a projected radius on the order of the impact parameter. When the impact parameter is on the order of the Einstein radius, this corresponds to the inner mass enclosed within that radius, and the lens's actual virial mass may be a few orders of magnitude larger $M_L \sim 10^3$--$10^6\,M_\odot$. Those mass scales are relevant for collapsed DM halos in CDM theories. Meanwhile, matter distribution on those scales may be smoothed out in alternative micro-models for the DM. However, observing sub-galactic DM clumps is in general difficult due to the lack of electromagnetic emission. Gravitational wave observations therefore offer a precious window into the matter distribution in the Universe on very small scales.

Lensing in the geometrical regime preserves the shape of the waveform. Without electromagnetic observation, it is difficult to disentangle between the true source distance and the lensing magnification. Inference about the lens therefore requires detecting multiple images. By contrast, wave diffraction induces amplitude and phase modulations in the frequency domain waveform. Those modulations are observable imprints of lensing even though multiple images do not always form --- for example in the case of a relatively large impact parameter or a shallow inner density profile of the lens. 

In the single-image regime, diffraction-induced modulations are small in size~\cite{Takahashi:2003ix}. Typically, the amplitude modulation is no larger than a few tens of percent in fraction, and the phase modulation is less than a few tens of percent of a radian, although those can be enhanced in the presence of an external shear. Since the overall distortion to the waveform is moderate, standard matched filtering using unlensed templates would yield a good match. Nevertheless, we will show that a diffraction-distorted waveform is indeed distinguishable from the unlensed waveform provided that the matched filtering signal-to-noise ratio (SNR) is sufficiently high (e.g. $\gtrsim 20$--$30$).

In previous studies, the detectability of the lensing diffraction effects was often estimated based on the technique of matched filtering~\cite{Takahashi:2003ix, 2014PhRvD..90f2003C, Jung:2017flg}, which requires specifying a lensed waveform model. For idealized lenses, such as point masses or singular isothermal spheres, it is feasible to construct a parametrized model for the mass profile and derive the corresponding diffraction signature. However, for realistic lenses this approach can be cumbersome due to the large number of parameters needed. Moreover, the correct lens profile to use may not be confidently known from theory or from simulations, especially for low mass DM halos.

Another issue overlooked in previous works was the look-elsewhere effect, which reduces detection significance. This is particularly pertinent because a large number of possible lensed waveforms need to be searched for, and because lensed events are expected to be rare. Any practical detection method must allow for correct quantification of the look-elsewhere effect.

To address the above issues, we present a new method based on dynamic programming. The method is computationally cheap and is highly practical as it does not require any parametrized model for lensed waveforms. The key idea is that diffraction-induced amplitude and phase distortions are highly correlated in the frequency domain, unlike the (nearly) stationary detector noise which has little correlation between different frequency components. We therefore compute a marginalized likelihood over all possible waveform perturbations around the best-fit unlensed waveform, assigning a prior probability for random amplitude and phase perturbations such that correlated perturbations are favored. Under the assumption of a Markovian process, this marginalized likelihood can be efficiently computed with the Forward algorithm~\cite{rabiner1989tutorial}. The false positive and the false negative probabilities can then be quantified through the Monte Carlo technique, which properly accounts for the look-elsewhere effect.

As a proof of concept, we will assess the observational prospect of our method applied to compact binary coalescence, using a pseudo-Jaffe lens with a characteristic mass $M_E \sim 10^2$--$10^3\,M_\odot$ enclosed within the Einstein radius. For a pair of fully upgraded aLIGO detectors, we find that the horizon distance of sensitivity for binary neutron star (NS) mergers is not likely to be promising in terms of probing a substantial amount of line-of-sight mass, but that for binary black hole (BH) mergers can reach as far as $\sim 1\,$Gpc (effective luminosity distance). The prospect will be further enhanced with joint detection by additional detectors in the network. As for one third-generation detector, such as the proposed Einstein Telescope (ET)~\cite{Hild:2008ng}, the horizon for suitable binary BH sources will be dramatically extended to $\gtrsim 10\,$Gpc, in which case the line of sight can have a significantly larger chance intersecting low mass halos.

The remainder of this paper is organized as the following. In \refsec{diff}, we review the physics of lensing in the wave diffraction regime, with emphasis on the general behaviors of the diffractive distortion. We then discuss how to detect diffraction signals in \refsec{detection}. We first develop intuition using the idealized method of matched filtering (\refsec{mf}). We then present a practical detection method based on dynamic programming (\refsec{dynamic}). In \refsec{sim}, we demonstrate the method of dynamic programming by performing mock detection with binary NS and BH mergers. Assuming a representative lens profile, we estimate detectability for those GW sources at second-generation detector networks and at third-generation detectors. In \refsec{issue}, we briefly discuss whether or not diffraction induced modulations may be degenerate with the effects of spin-orbit precession and orbital eccentricity on the waveform. Finally, we present summarizing discussion in \refsec{concl}. 

\section{Diffraction distortion in waveforms}
\label{sec:diff}

Consider a lens at redshift $z_L$ and a GW source at redshift $z_S$ in a flat Friedmann-Lema\^{i}tre-Robertson-Walker universe. Let $d_L$, $d_S$ and $d_{LS}$ be the angular diameter distances to the lens, to the source, and from the lens to the source, respectively. At any given observed frequency $f$, the lensed waveform is $h(f) = F(f)\,h_0(f)$, where $h_0(f)$ is the unlensed waveform. Under the approximation of a single mass sheet, the multiplicative factor $F(f)$ is a complex number and can be obtained from an diffraction integral~\citep{schneider1992gravitational}
\ba
\label{eq:Fw}
\hspace{-0.5cm}
F(f) = \frac{f\,(1+z_L)}{i}\,\frac{d_L\,d_S}{c\,d_{LS}}\,\int\,d^2\bfx\,e^{ i\,2\pi\,f\,(1+z_L)\,\tau(\bfx)},
\ea
where $\bfx$ are the angular coordinates on the lens plane. The ray travel time $\tau(\bfx)$, defined relative to free propagation, can be written as the sum of the geometrical delay term and the Shapiro delay term, $\tau(\bfx)= (d_L\,d_S)/(c\,d_{LS})\,\left( \bfx\cdot\bfJ_{\rm ext}\cdot\bfx/2 - \phi(\bfx) \right)$, where $\phi(\bfx)$ is the lensing potential, and we introduce a Jacobian matrix $\bfJ_{\rm ext}$ to account for any possible external convergence and shear.

At high frequencies, namely in the geometrical limit, $F(f)$ is the sum of contributions from one or multiple images, which we label as $a = 1, 2, \cdots$~\citep{schneider1992gravitational}, 
\ba
F_{\rm geo}(f) = \sum_a\,\sqrt{|\mu(\bfx_a)|}\,e^{-i\,\pi\,\delta_a\,{\rm sgn}(f)}\,e^{i\,2\pi\,f\,(1+z_L)\,\tau(\bfx_a)}.
\ea
At each image position $\bfx_a$, $\mu(\bfx_a)$ is the signed magnification factor. The summation over $i$ accounts for the possibility of multiple images. The Morse phase $e^{-i\,\pi\,\delta_i}$~\citep{ambrose1961index} depends on the image type and represents a residual wave effect of topological origin~\citep{Dai:2017huk}. 

In the geometrical limit, waveform distortions can only arise when multiple images mutually interfere. The existence of more than one image often requires a sufficiently compact lens and a small impact parameter. If only one image is present $\bfx_a = \bfx_I$, the lensed waveform is a rescaled version of the intrinsic waveform but is shifted by $\tau(\bfx_1)$ in the time domain. In this case, a lensed event is indistinguishable from an unlensed one, unless either the luminosity distance or the source redshift is independently measured~\citep{Dai:2016igl, Dai:2017huk}.

In the absence of multiple-image interference, the measurable effect of lensing is encoded in the deviation of $F(f)$ from $F_{\rm geo}(f)$,
\ba
F_{\rm rel}(f) := F(f)/F_{\rm geo}(f),
\ea
which induces waveform distortions. By construction, $F_{\rm rel}(f)$ approaches unity in the limit of high frequencies $f \rightarrow \infty$.

We now study concrete examples by modeling the possible intervening lenses using pseudo-Jaffe ellipsoids~\citep{jaffe1983simple}. We first define the Einstein angular radius $\theta_E := 4\pi\,(\sigma_v/c)^2\,(d_{LS}/d_S)$. The effective velocity dispersion $\sigma_v$ is related to the characteristic lens mass, defined to be the enclosed mass within the Einstein radius:
\ba
M_E & = & \left( 4\pi^2\,\sigma^4_v\,d_{\rm eff}\right)/\left(G\,c^2\right) \\
& = & 100\,M_\odot\,\left( \frac{\sigma_v}{1\,{\rm km/s}} \right)^4\,\left( \frac{d_{\rm eff}}{1\,{\rm Gpc}} \right). \nonumber
\ea
where $d_{\rm eff} := d_L\,d_{LS}/d_S$. The convergence is given by
\ba
\kappa = (\theta_E/2)\,\left[ \left(s^2 + \xi^2\right)^{-1/2} - \left(a^2 + \xi^2\right)^{-1/2} \right],
\ea
Here $s$ is the core scale and $a$ is the truncation scale. The ellipse variable $\xi$ is introduced to allow for ellipticity. In a coordinate system where the major axes of the lens ellipse align with the coordinate axes, we have $\xi^2 = x^2_1 + x^2_2/q^2$ for $0 < q \leqslant 1$. The case $q=1$ corresponds to an axisymmetric lens. Analytic results for the lensing potential $\phi(\bfx)$ can be found in Ref.\cite{Keeton:2001ss}.

We choose this simple analytic lens model because it can approximate reasonably well any virialized self-gravitating mass clump with an inner core and an outer radius of truncation.

The importance of diffraction effects is characterized by a dimensionless parameter
\ba
\label{eq:wpar}
w & := & 2\pi\,f\,(1+z_L)\,\frac{d_L\,d_S}{c\,d_{LS}}\,\theta^2_E \\
& \simeq & 1.3\,(1+z_L)\,\left( \frac{f}{10^2\,{\rm Hz}} \right)\,\left( \frac{\sigma_v}{1\,{\rm km/s}} \right)^4\,\left( \frac{d_{\rm eff}}{1\,{\rm Gpc}} \right).
\ea
It is linearly proportional to $M_E$ at fixed wave frequency. 

\begin{figure}[h]
  \begin{center}
    \includegraphics[scale=0.6]{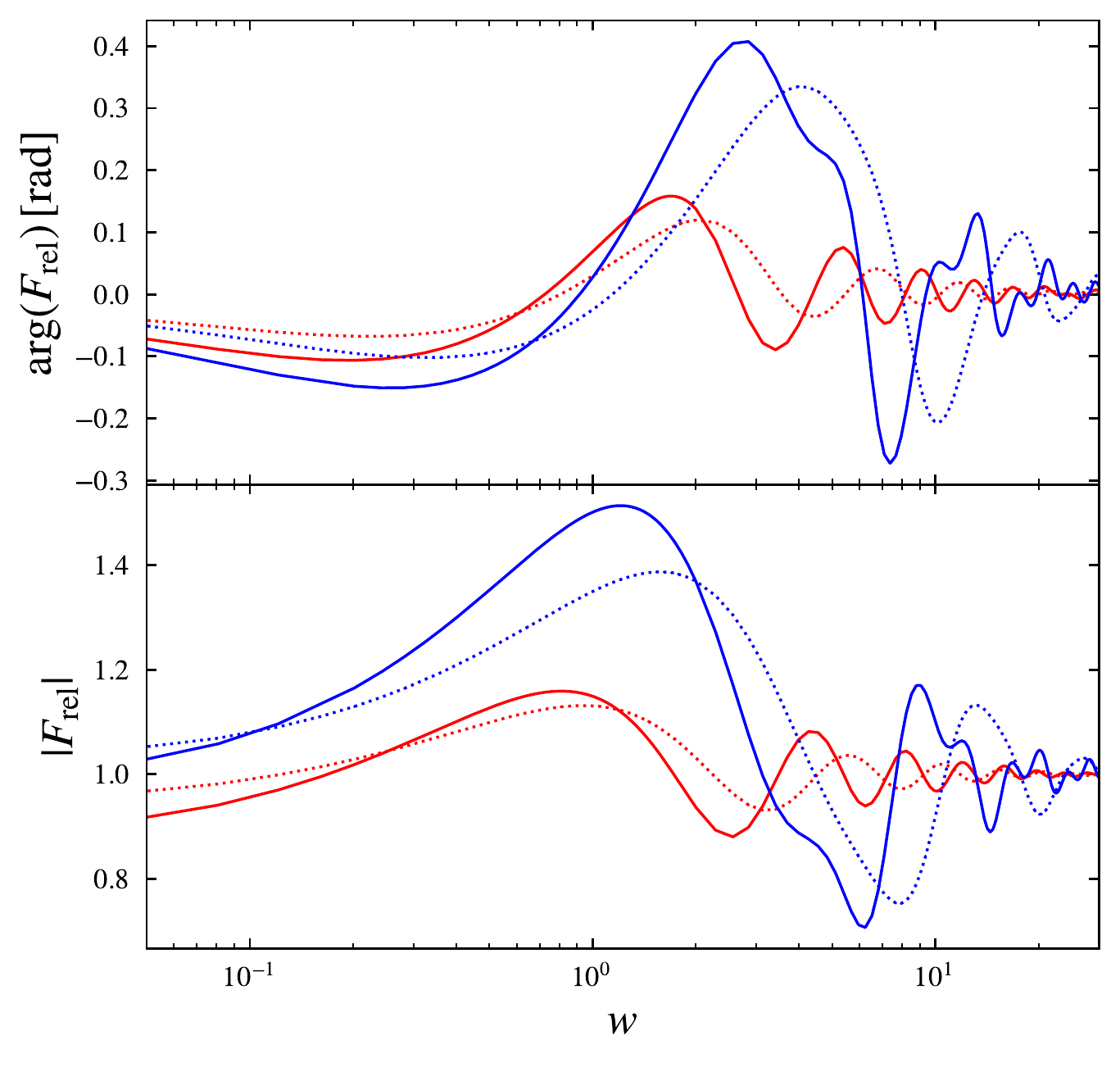}
    \caption{\label{fig:Fwrel_example} Examples of the relative amplification factor $F_{\rm rel}(w)$ for pseudo-Jaffe ellipsoids. We assume $s = 0.1$ and $a = 2$, and an impact parameter $\bfy = [0.8,\,0.8]$. Four cases are shown: (1) axisymmetric, no external convergence/shear (solid red); (2) $q = 0.5$, no external convergence/shear (dotted red); (3) axisymmetric, $\kappa_{\rm ext} = \gamma_{\rm ext} = 1/3$ (solid blue); (4) $q = 0.5$, $\kappa_{\rm ext} = \gamma_{\rm ext} = 1/3$ (dotted blue). When both the lens ellipticity and the external shear are non-zero, we assume a misalignment angle $\pi/2$ between their major axes. (All angular variables are in units of $\theta_E$.)}
  \end{center}
\end{figure}

\begin{figure}[h]
  \begin{center}
    \includegraphics[scale=0.6]{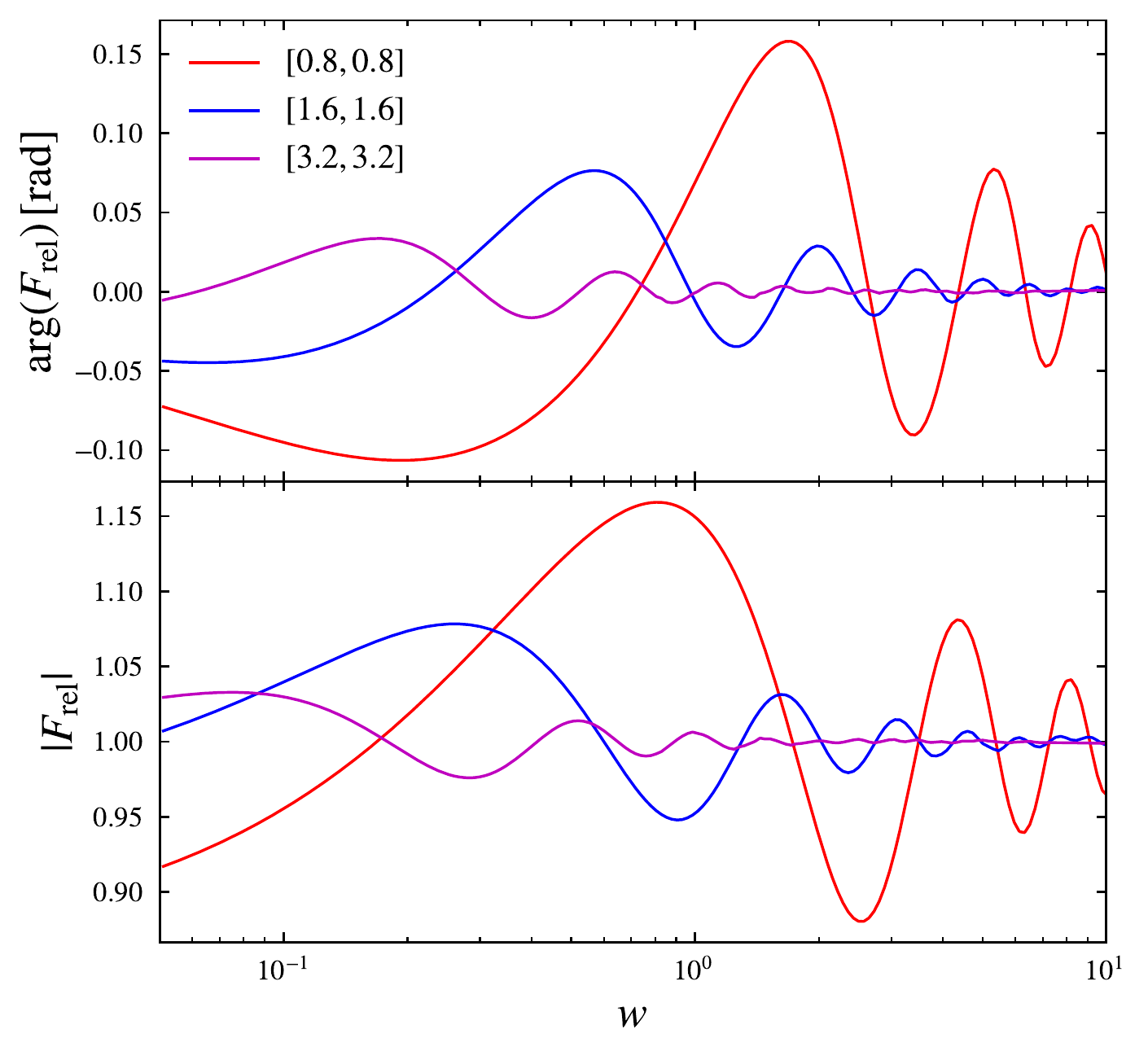}
    \caption{\label{fig:Fwrel_impact_parameter} Same as \reffig{Fwrel_example} but for the case of an axisymmetric lens $a = 2$ and $s = 0.1$ without any external convergence/shear. Various curves correspond to different source impact parameters $\bfy$, whose values are indicated in the legends.}
  \end{center}
\end{figure}

\reffig{Fwrel_example} shows examples of $F_{\rm rel}(w)$. Typically, $F_{\rm rel}(w)$ asymptotes to unity for $w \gg 10$ if the number of geometrical image is one.  For $w \lesssim 10$, amplitude and phase modulations become non-negligible but do not exceed $\sim 10\%$--$20\%$. The modulations can be enhanced in the presence of order-unity external convergence $\kappa_{\rm ext}$ and shear $\gamma_{\rm ext}$, a situation that may arise if the lens is embedded in a larger lens (e.g. lensing by a subhalo residing in the halo of an intervening galaxy lens). Among the many modulation cycles, the first one typically has the largest size and should be the most interesting for detection.

\reffig{Fwrel_impact_parameter} shows how $F_{\rm rel}(w)$ depends on the impact parameter $\bfy$ for an isolated axi-symmetric lens $s=0.1$ and $a = 2$. The sizes of both the phase and the amplitude modulations decrease as the inverse of $|\bfy|$. Also, the locations of the maxima and the minima in terms of $w$ scale as the inverse of $|\bfy|$. This implies that at fixed physical frequency $f$ and distance $d_{\rm eff}$ and for the same lens, a lensing configuration with a larger impact parameter is sensitive to a smaller mass $M_E$ enclosed within the Einstein radius.

\begin{figure}[t]
  \begin{center}
    \includegraphics[scale=0.6]{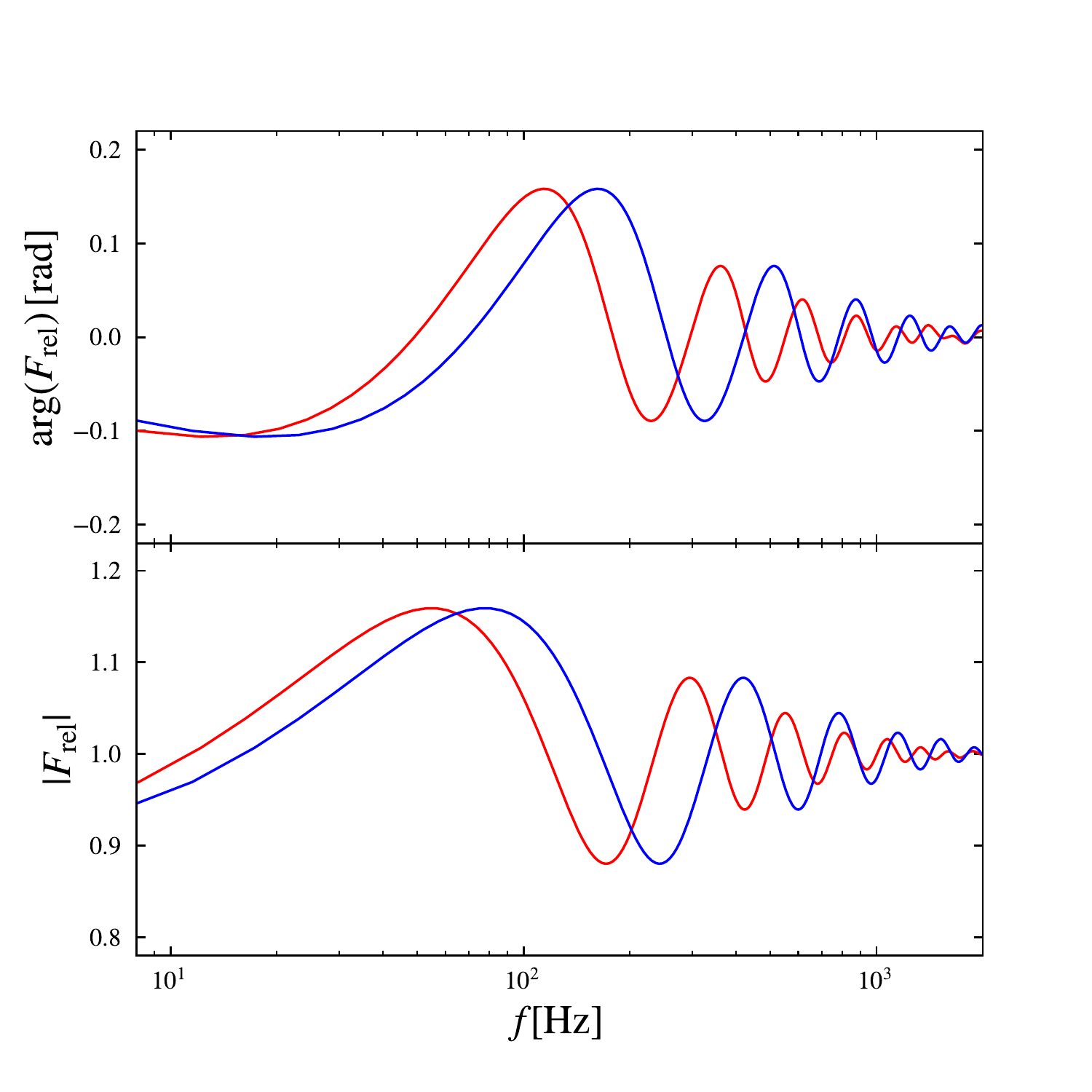}
    \caption{\label{fig:f_Fwrel_example} Same as \reffig{Fwrel_example} but mapped to wave frequencies $f$ in the LIGO band in physical units. Curves are calculated for an axisymmetric lens with $s = 0.1$, $a = 2$ and $\bfy = [0.8,\,0.8]$ without external convergence/shear. Two cases are shown: (1) $\sigma_v = 2.0\,{\rm km/s}$, $z_L = 0.04$, and $d_{\rm eff} = 70\,{\rm Mpc}$ (red); (2) $\sigma_v = 1.2\,{\rm km/s}$, $z_L = 0.2$, and $d_{\rm eff} = 330\,{\rm Mpc}$ (blue).}
  \end{center}
\end{figure}

\reffig{f_Fwrel_example} plots $F_{\rm rel}(f)$ in the frequency band of ground-based detectors. Detectable lenses should have $\sigma_v$ and $d_{\rm eff}$ in the ``sweet spot'' such that the ground-based frequency band maps to $w \sim \mathcal{O}(1)$. For instance, an binary NS merger event from $z_S = 0.07$ with a luminosity distance $D \simeq 300\,{\rm Mpc}$ can be sensitive to a pseudo-Jaffe lenses with a velocity dispersion $\sigma_v \simeq 2\,{\rm km/s}$ at $z_L = 0.04$ ($d_{\rm eff} \approx 70\,{\rm Mpc}$). This translates to an intriguingly small Einstein mass $M_E \approx 100\,M_\odot$. The more massive binary BH mergers are detectable out to larger distances. A binary BH event from $z_S = 0.4$ with a luminosity distance $D \simeq 2\,{\rm Gpc}$ can probe lenses with $\sigma_v \sim 1\,{\rm km/s}$ and $M_E \sim 70\,M_\odot$. The nearly unique order of magnitude in $M_E$ is set by the wave frequency in the detector's band, whose inverse should match the Schwarzschild time scale $\sim G\,M_E/c^3$ in order to maximize the diffraction effects. 

\section{Detection of diffraction effects}
\label{sec:detection}

In this Section, we discuss the detectability of diffraction-induced modulations in the waveform. 

\subsection{Detection by matched filtering}
\label{sec:mf}

\begin{figure}
  \begin{center}
    \includegraphics[scale=0.6]{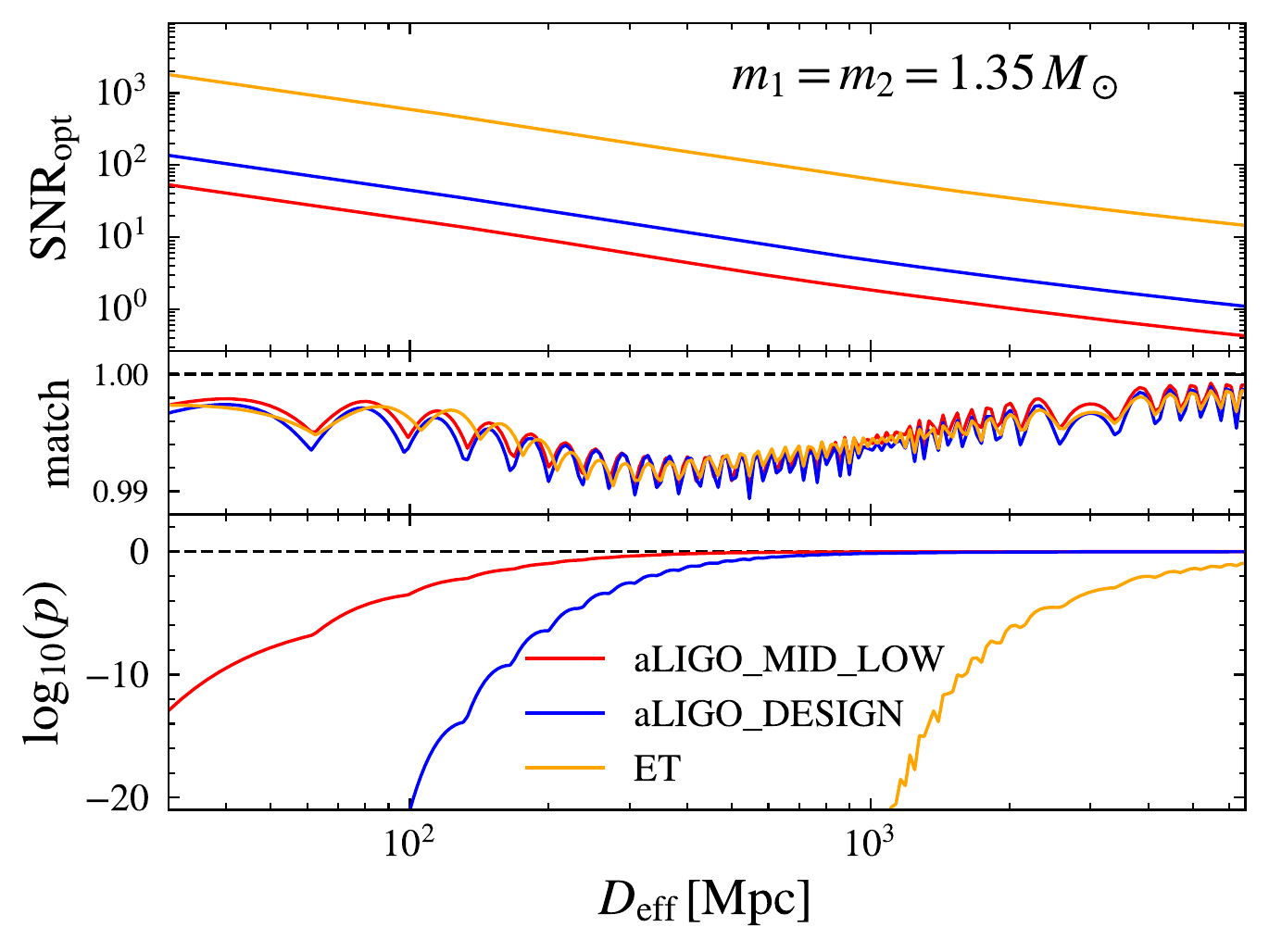}\\
    \vspace{-0.2cm}\includegraphics[scale=0.6]{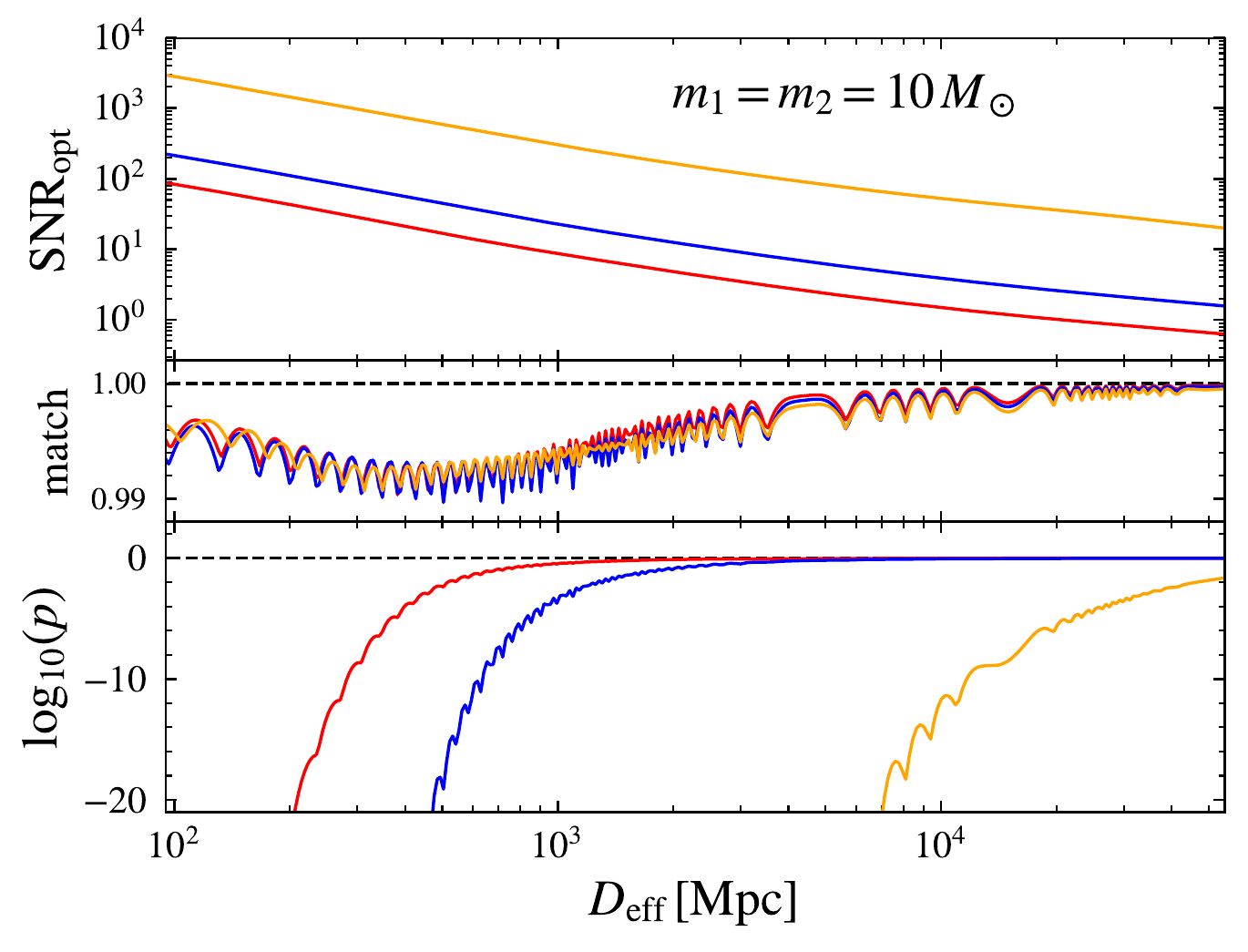}\\
    \vspace{-0.2cm}\hspace{0.1cm}\includegraphics[scale=0.6]{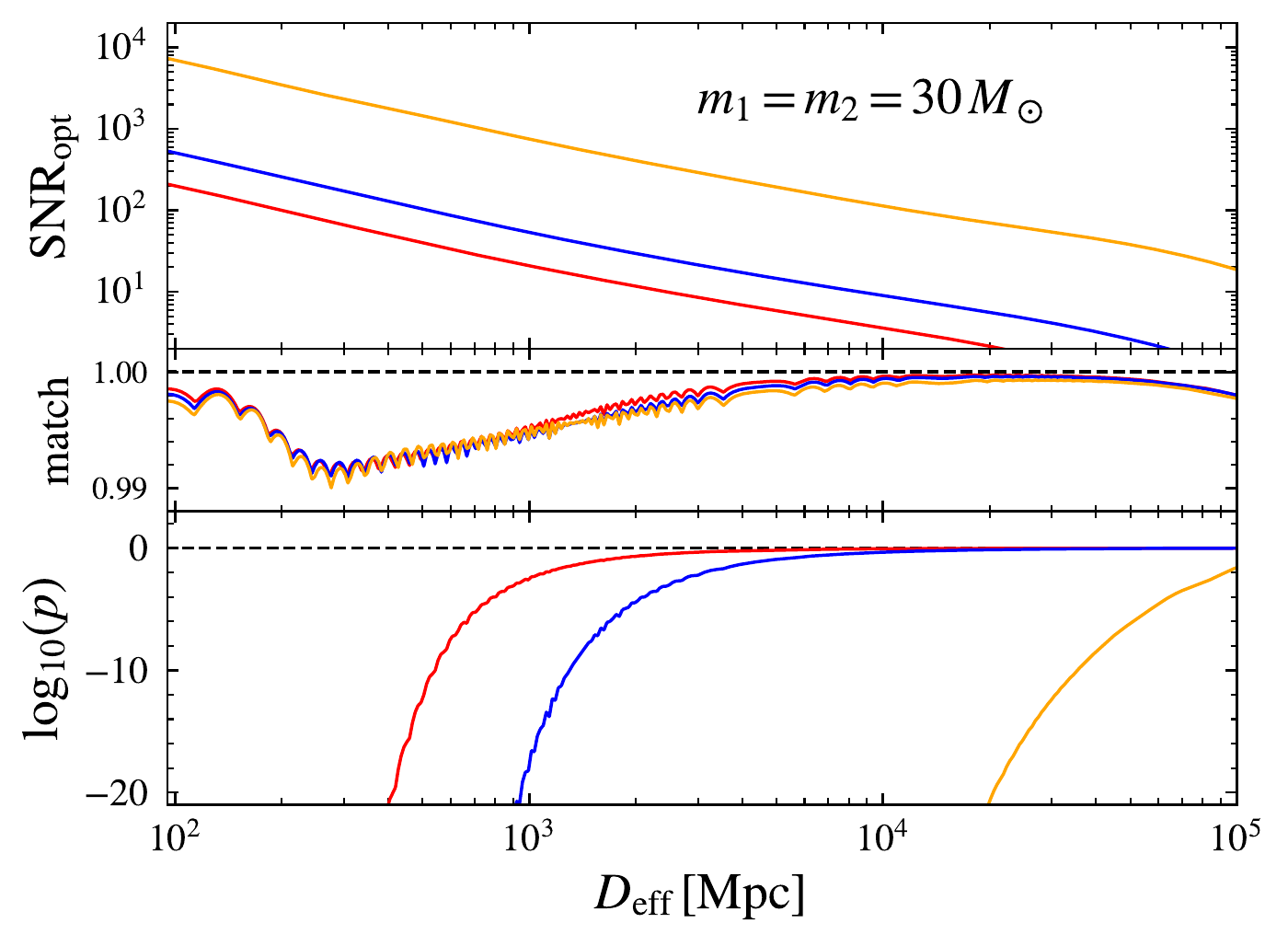}
    \caption{\label{fig:snr_Deff} Examples for binary NS (top), $10\,M_\odot$ binary BH (middle), and $30\,M_\odot$ binary BH (bottom). Non-spinning waveforms are injected. In each plot, we show the optimal matched filtering SNR (upper panel), the ``match'' between the unlensed waveform $h_0(f)$ and the lensed waveform $h_L(f)$ quantified as $\left|\VEVZ{h_0|h_L}\right|/\sqrt{\VEV{h_L|h_L}\,\VEV{h_0|h_0}}$ (middle panel), and a corresponding $p$-value (c.f. \refeq{lnp}), all as a function of $D_{\rm eff}$. We compute for three noise PSDs:\texttt{aLIGO\_MID\_LOW} (red), \texttt{aLIGO\_DESIGN} (blue), and the proposed ET~\cite{Hild:2008ng} (orange). The aLIGO sensitivity curves are provided in \texttt{LALSuite}.  All curves are computed for a single detector and a frequency range $f\in [10,\,1024]\,$Hz. Refer to the text for more information.}
  \end{center}
\end{figure}

The ideal method is to construct waveform templates that incorporate the exact amplitude and phase modulations, and to perform a matched-filtering search using those templates. The significance of the matched-filtering method is quantified by the SNR. At a single detector, the strain time series $s(t) = h(t) + n(t)$ is the sum of the GW signal $h(t)$ and the detector noise $n(t)$. For a waveform template $h_T(t)$ defined up to an arbitrary normalization $\lambda$ and an arbitrary phase constant $\phi_c$, the matched-filtering SNR has a maximal value
\ba
\label{eq:SNRperdet}
{\rm SNR}^2 & = & \max_{\lambda,\, \phi_c}\,\left[ \VEV{s - \lambda\,e^{i\phi_c}\,h_T|s - \lambda\,e^{i\phi_c}\,h_T}  -\VEV{s|s} \right] \en
& = & \left|\VEVZ{s| h_T}\right|^2 / \VEV{h_T| h_T},
\ea
with a best-fit normalization $\lambda = |\VEVZ{s|h_T}|/\VEV{h_T|h_T}$. Here $\VEV{a|b}$ denotes the ``overlap'' between any two strain series $a(t)$ and $b(t)$, and has the following frequency-space representation,
$\VEV{a|b} := 4\,{\mathfrak Re}\,\int^{+\infty}_0\,df\,a(f)\,b^*(f)/S_N(f)$, where $S_N(f)$ is the one-sided power spectrum density (PSD) for the detector noise (assumed to be Gaussian). While $\VEV{a|b}$ is always real, we also introduce a complex-valued ``overlap'' $\VEVZ{a|b} := 4\,\int^{+\infty}_0\,df\,a(f)\,b^*(f)/S_N(f)$, which is a useful quantity to compute when one would like to vary the phase constant $\phi_c$ in order to maximize the match. 

A GW event may be simultaneously seen at multiple detectors. Strictly speaking, the waveform normalization, the phase constant, and the arrival time are all correlated between the detectors, depending on the source's sky coordinates and the detectors' locations and orientations. Since those information is not our focus here, we neglect those correlations for simplicity~\cite{Dai:2018dca, Roulet:2018jbe}. In this case, the overall SNR is given by the SNRs defined in \refeq{SNRperdet} for individual detectors added up in quadrature.

In the presence of lensing  $h_L(f) = F(f)\,h_0(f) = F_{\rm rel}(f)\,F_{\rm geo}(f)\,h_0(f)$, we have $s(f) = h_L(f)  +n(f)$. If the exact diffraction-distorted waveform $h_L(f)$ is used as the template, the optimal matched-filtering SNR is
\ba
{\rm SNR}^2_{\rm opt}= \left|\VEVZ{s|h_L}\right|^2/\VEV{h_L|h_L} \approx \VEV{h_L|h_L},
\ea
where we have neglected the overlap between $h_L(f)$ and $n(f)$. However, lensed GW signal can also be recovered with an unlensed template, say using $h_{\rm geo}(f) := F_{\rm geo}(f)\,h_0(f)$, albeit at a reduced SNR. This is because the phase distortion in $F_{\rm rel}(f)$ is typically much less than one radian. The SNR corresponding to the unlensed template is
\ba
{\rm SNR}^2_{\rm unlen} = \frac{\left|\VEVZ{s | \tilde h_{\rm geo}}\right|^2}{\VEV{\tilde h_{\rm geo} | \tilde h_{\rm geo}}} \approx \frac{\left|\VEVZ{h_L|\tilde h_{\rm geo}}\right|^2}{\VEV{\tilde h_{\rm geo} | \tilde h_{\rm geo}}} = \frac{\left|\VEVZ{h_L| h_{\rm BF}}\right|^2}{\VEV{ h_{\rm BF} |  h_{\rm BF}}}.
\ea
The tilde added to $h_{\rm geo}(f)$ is a notation for enumerating all possible values of $t_c$ to $h_{\rm geo}(f)$ in order to maximize the match. The best-fit (unlensed) template
\ba
\label{eq:hBF}
h_{\rm BF}(f) = \frac{\left|\VEVZ{h_L|\tilde h_{\rm geo}}\right|}{\VEV{\tilde h_{\rm geo}|\tilde h_{\rm geo}}}\,\tilde h_{\rm geo}(f)\,e^{i\,{\rm arg} \VEVZ{h_L|\tilde h_{\rm geo}}}.
\ea
Intuitively, using the correct template generally yields a better match, since ${\rm SNR}^2_{\rm opt} - {\rm SNR}^2_{\rm unlen} > 0$ due to the Cauchy-Schwarz inequality.

How statistically significant is the improvement in the SNR by using the lensed template relative to using the unlensed one? We would like to define a $p$-value which quantifies the chance that there are no amplitude and phase modulations and the SNR improves due to a statistical fluke. One definition would be the change in the likelihood (per detector)
\ba
\label{eq:lnp}
\ln p & = & -  \left( {\rm SNR}^2_{\rm opt} - {\rm SNR}^2_{\rm unlen}  \right)/2  \en
& \approx & - \frac12\,\left( \VEV{h_L|h_L} - \frac{\left|\VEVZ{h_L| h_{\rm BF}}\right|^2}{\VEV{ h_{\rm BF} |  h_{\rm BF}}} \right).
\ea
Ref.~\cite{Jung:2017flg} instead uses the vector-space ``distance''
\ba
\label{eq:lnphat}
\ln p  = -  \langle h_L - h_{\rm BF}| h_L - h_{\rm BF} \rangle / 2.
\ea
\refeq{lnp} and \refeq{lnphat} are equivalent as long as $h_{\rm BF}(f)$ has the best-fit normalization and is tuned to the best-fit phase constant as in \refeq{hBF}.

In \reffig{snr_Deff}, we estimate how well the lensed waveform can be distinguished from the unlensed waveform depending on the source distance. We consider a specific lens: a pseudo-Jaffe sphere with $\sigma_v = 2\,{\rm km/s}$ located at $z_l = z_s/2$, having the parameters of Case (1) in \reffig{Fwrel_example}. The curves are computed for the optimal source location and orientation for which $D_{\rm eff}$ is equal to the luminosity distance. The lensed and the unlensed waveforms always have a good match (better than 99\%), reflecting the small sizes of amplitude and phase modulations. Nevertheless, it is possible to extract the subtle difference through matched filtering if the precise lensed waveform is known.

With a single aLIGO detector at the design sensitivity (\texttt{aLIGO\_DESIGN}), the diffraction signature should be detectable (say require $p < 10^{-6}$) for binary neutron stars within $D_{\rm eff} \approx 200\,$Mpc, and for heavy binary BHs ($30\,M_\odot$ per component) within $D_{\rm eff} \approx 3\,$Gpc. For a future detector of the third generation, these distances increase to $1\,$Gpc and $50\,$Gpc, respectively. 

In the case of joint detection with a network of $N_{\rm det}$ detectors of comparable sensitivity, since the same modulation is imprinted at all detectors, the logarithm of the $p$-value is multiplied by a factor of $N_{\rm det}$, which results in further increase in the horizon distance. For the same detection significance, two identical aLIGO detectors at the design sensitivity could jointly reach $D_{\rm eff} \sim 300\,$Mpc for binary NS mergers and $D_{\rm eff} \sim 2\,$Gpc for 30-solar-mass binary BH mergers.

Admittedly, \reffig{snr_Deff} overestimates the detectability. First, diffraction-induced modulations are partially degenerate with changes in the intrinsic parameters (e.g. chirp mass, mass ratio, spins, tidal deformabilities, etc.). Moreover, one needs to account for the look-elsewhere effect when enumerating a large number of possible modulations. Still, \refeq{lnp} provides zeroth-order intuition toward understanding this problem. In the following, we address these issues by developing a practical detection method using dynamic programming.

\subsection{Detection by dynamic programming}
\label{sec:dynamic}

The matched filtering method requires a template for $F_{\rm rel}(f)$. However, the exact shape of $F_{\rm rel}(f)$ will not be known {\it a priori}. It depends on many unknown parameters including the lens mass, the lens mass profile and shape, and the impact parameter. 

One strategy is to perform an agnostic search for all possible functional forms $F_{\rm rel}(f) = g(f)$. Define the following score,
\ba
\label{eq:mcalS}
\calS := \int\,\mathcal{D} g(f)\,\calP\left[ g(f) \right]\,\prod^{N_d}_{a=1}\,\frac{P[ s_a(f) | g(f)\,h_{{\rm BF},a}(f) ]}{P[ s_a(f) | h_{{\rm BF},a}(f) ]},
\ea
This is a ``path integral'' over all possible amplitude and phase distortions $g(f)$. In the numerator, we explore perturbations to the {\it best-fit} unlensed waveform, $g(f)\,h_{{\rm BF},a}(f)$, enumerating all detectors $a = 1,2,\cdots, N_d$. Here $\calP[g(f)]$ is the prior probability for any specific $g(f)$. The notation $P[s_a(f)|h_a(f)]$ denotes the matched filtering likelihood for the strain data $s_a(f)$ given a putative GW signal $h_a(f)$ at the $a$-th detector. \refeq{mcalS} measures the marginalized improvement in the likelihood when the best-fit unlensed waveform is perturbed by appropriate amounts.

Random $g(f)$ can happen to improve the match due to detector noise. However, stationary Gaussian noise has zero correlations between frequencies. By contrast, diffraction induces amplitude and phase modulations that are correlated between frequencies, as can be seen from \reffig{Fwrel_example} and \reffig{f_Fwrel_example}. In other words, detector noise matters more for rapidly oscillating realizations of $g(f)$, while diffraction corresponds to a continuous and smooth $g(f)$. Therefore, the diffraction signature is distinguishable from random noise if one uses a suitable prior $\calP[g(f)]$ that favors continuous and smooth functional forms.

In practice, the functional integral of \refeq{mcalS} can be approximated by a summation over a discrete set of $g(f)$'s. Consider $N$ frequency bins $f_j \leqslant f < f_{j+1}$, labelled by $j  = 0, 1, \cdots, N-1$. We can approximate a continuous function $g(f)$ with a series of ``steps''
\ba
g(f) = \sum^{N-1}_{j=0}\,\left( 1 + u_j + i\,v_j \right)\,\Theta\left( f - f_j \right)\,\Theta\left( f_{j+1} - f \right),
\ea
where $\Theta(x)$ is the usual Heaviside function, and $u_j$ and $v_j$ are fractional perturbations to the real part and the imaginary part, respectively. A discretized $g(f)$ is specified by a set of coefficients $\{\bfu,\,\bfv\}:=\{u_j,\,v_j\}$ for $j=0,1,\cdots,N-1$, which we assume take discrete values within some range. For example, we may allow them to take values on uniform grids:
\ba
\label{eq:uvgrids}
&& u_j \in \{ u_{\rm min} + k\,(u_{\rm max}  - u_{\rm min})/n_u,\, k  = 0, 1, \cdots, n_u \} \en
&& v_j \in \{ v_{\rm min} + l\,(v_{\rm max}  - v_{\rm min})/n_v,\, l  = 0, 1, \cdots, n_v \} \en
\ea
Then \refeq{mcalS} can be approximated as
\ba
\label{eq:mcalSdscr}
&& \calS = \left[ \prod^{N-1}_{j = 0} \left( \sum_{u_j}\,\sum_{v_j} \right) \right]\,\calP\left[ \{\bfu,\, \bfv\} \right]\en
&& \times \prod^{N-1}_{j=0}\,\left( \prod^{N_d}_{a = 1}\,\frac{P_j[s_a(f) | h_{{\rm BF}, a}(f)\,(1+u_j + i\,v_j)]}{P_j[s_a(f)| h_{{\rm BF}, a}(f)]} \right).\quad
\ea
The logarithm of the relative likelihood associated with the $j$-th frequency bin is given by
\ba
&& \ln \frac{P_j[s_a(f) | h_{{\rm BF}, a}(f)\,(1+u_j + i\,v_j)]}{P_j[s_a(f)| h_{{\rm BF}, a}(f)]} \en
& = & \VEV{s_a | h_{{\rm BF},a}\,(1+u_j + i\,v_j)}_j - \VEV{s_a| h_{{\rm BF}, a}}_j \en
&& - \frac12\,\VEV{h_{{\rm BF},a}\,(1+u_j + i\,v_j)|h_{{\rm BF},a}\,(1+u_j + i\,v_j)}_j \en
&& + \frac12\,\VEV{h_{{\rm BF}, a}|h_{{\rm BF}, a}}_j,
\ea
where we introduce the notation $\VEV{s_a(f)|h_a(f)}_j := 4\,\mathfrak{Re}\,\int^{f_{j+1}}_{f_j}\,df\,s_a(f)\,h_a^*(f)/S_{N,a}(f)$ for the $j$-th frequency bin at the $a$-th detector

The prior function $\calP[\{\bfu,\,\bfv\}]$ remains to be specified. We assume that the (discretized) $g(f)$ can be viewed as a Markovian process in frequency space, so that $\calP[\{\bfu,\,\bfv\}]$ recursively factorizes following the chain rule of conditional probability:
\ba
\calP[\{\bfu, \bfv\}] = \calP[u_0, v_0]\,\prod^{N-1}_{j=1}\calP[u_j, v_j|u_{j-1}, v_{j-1}].
\ea
In this case, \refeq{mcalSdscr} can be efficiently computed by the Forward algorithm.

Let us order the frequency bins from the lowest to the highest. Imagine the ``path integral'' of \refeq{mcalSdscr} is only performed for the first $n+1 \leqslant N$ frequency bins $j = 0, 1, \cdots, n$. Define the following ``partial'' path integral
\ba
\hspace{-0.7cm}
&& \calS_n(u_n, v_n) =  \left[ \prod^{n-1}_{j = 0} \left( \sum_{u_j}\,\sum_{v_j} \right) \right] \,\Bigg\{ \left( \prod^{n-1}_{j=0} \calP\left[ u_j, v_j | u_{j-1}, v_{j-1} \right]\right)   \en
\hspace{-0.7cm}
&& \times \prod^{n-1}_{j=0}\,\left( \prod^{N_d}_{a = 1}\,\frac{P_j[s_a(f) | h_{{\rm BF}, a}(f)\,(1+u_j + i\,v_j)]}{P_j[s_a(f)| h_{{\rm BF}, a}(f)]} \right).\en
\hspace{-0.7cm}
&& \times \calP[u_n, v_n |u_{n-1}, v_{n-1}] \Bigg\} \en
\hspace{-0.7cm}
&& \times \left( \prod^{N_d}_{a = 1}\, \frac{P_n[s_a(f) | h_{{\rm BF}, a}(f)\,(1+u_n + i\,v_n)]}{P_n[s_a(f)| h_{{\rm BF}, a}(f)]} \right).
\ea
This leads to a recursive algorithm with a polynomial computational cost $\mathcal{O}(n_u\,n_v\,N)$:
\ba
&& \calS_n(u_n, v_n) = \left( \prod^{N_d}_{a = 1}\, \frac{P_n[s_a(f) | h_{{\rm BF}, a}(f)\,(1+u_n + i\,v_n)]}{P_n[s_a(f)| h_{{\rm BF}, a}(f)]} \right) \en
&& \times \sum_{u_{n-1}}\,\sum_{v_{n-1}}\,\calP[u_n, v_n |u_{n-1}, v_{n-1}]\,\calS_{n-1}(u_{n-1}, v_{n-1}),
\ea
with initial conditions $\calS_{-1}(u_{-1}, v_{-1})\equiv 1$ and $\calP[u_0, v_0|u_{-1}, v_{-1}] \equiv 1$. The marginalized score of \refeq{mcalSdscr} is then given by
\ba
\calS = \sum_{u_{N-1}}\,\sum_{v_{N-1}}\,\calS_{N-1}(u_{N-1}, v_{N-1}).
\ea

To find out the best-fit ``path'', namely a most probable set of values $\{u_j=\hat u_j, v_j=\hat v_j\}$, we apply the Viterbi algorithm~\cite{viterbi1967error}. Let us define $\calV_j(u_j, v_j)$, which satisfies another recursion relation
\ba
&& \calV_n(u_n, v_n) = \left( \prod^{N_d}_{a = 1}\, \frac{P_n[s_a(f) | h_{{\rm BF}, a}(f)\,(1+u_n + i\,v_n)]}{P_n[s_a(f)| h_{{\rm BF}, a}(f)]} \right) \en
&& \times \max_{u_{n-1}, v_{n-1}}\,\calP[u_n, v_n |u_{n-1}, v_{n-1}]\,\calV_{n-1}(u_{n-1}, v_{n-1}),
\ea
with initial conditions $\calV_{-1}(u_{-1}, v_{-1})\equiv 1$ and $\calP[u_0, v_0|u_{-1}, v_{-1}] \equiv 1$. The ``end point'' of the most probable ``path'' is
\ba
\left( \hat u_{N-1},\,\hat v_{N-1} \right) = \argmax_{u_{N-1}, v_{N-1}}\,\calV_{N-1}(u_{N-1}, v_{N-1}).
\ea
One then traces backward: if for the $(j+1)$-th frequency bin $(\hat u_{j+1}, \hat v_{j+1})$ have been found, for the $j$-th frequency bin the best-fit ``path'' is
\ba
\left( \hat u_j,\,\hat v_j \right) = \argmax_{u_j, v_j}\,\calP[\hat u_{j+1}, \hat v_{j+1} |u_j, v_j]\,\calV_j(u_j, v_j).
\ea
This procedure then recovers the best-fit ``path'' $\{\hat u_j, \hat v_j\}$ for $j  = 0, 1, \cdots, N-1$.

The Markovian conditional probability $\calP[u_j, v_j| u_{j-1}, v_{j-1}]$ remains to be specified. To distinguish between diffraction and random noise, it should favor smooth ``paths''. For $\{u_j, v_j\}$ defined on uniform grids (\refeq{uvgrids}), a simple choice would be to require that from one frequency bin to the next the $u$-/$v$-coefficients may only ``jump'' by up to a maximum number of grid points. To be precise, let $u_j  = u_{\rm min} + k_j\,(u_{\rm max} - u_{\rm min})/n_u$, $v_j  = v_{\rm min} + l_j\,(v_{\rm max} - v_{\rm min})/n_v$, and define $\calP[k_j, l_j| k_{j-1}, l_{j-1}]:=\calP[u_j, v_j|u_{j-1}, v_{j-1}]$. We set $\calP[k_j, l_j| k_{j-1}, l_{j-1}]$ to be a nonzero constant if $ |k_j - k_{j-1}| \leqslant \Delta k_{\rm max}$ and $|l_j - l_{j-1}| \leqslant \Delta l_{\rm max}$ but otherwise zero, with the normalization $\sum^{n_u}_{k_j=0}\,\sum^{n_v}_{l_j=0}\,\calP[k_j, l_j| k_{j-1}, l_{j-1}] \equiv 1$.

Despite our approximation of $g(f)$ using a sequence of steps, the formalism can be straightforwardly generalized to more sophisticated models of $g(f)$. For instance, $g(f)$ can be approximated as linear or higher-order interpolation within frequency bins. Also, there is freedom to tune the specific form of $\calP[g(f)]$ under the Markovian assumption. In the following, we shall adopt the simplest scheme.

\section{Testing dynamic programming}
\label{sec:sim}

In this Section, we demonstrate dynamic programming as outlined in \refsec{dynamic} using mock GW signals. We apply the method of relative binning~\cite{Zackay:2018qdy, Dai:2018dca} for fast likelihood evaluations.

\subsection{Waveform models}
\label{sec:wfmodel}

For binary BHs, we use the phenomenological frequency-domain waveform model \texttt{IMRPhenomD}~\cite{Husa:2015iqa, Khan:2015jqa}. This model is applicable to inspiral, merger and ringdown of binary BHs with non-precessing spins.

In the frequency domain, the unlensed waveform can be written as $h_0(f) = A(f)\,e^{i\,\Psi(f)}$, where $A(f)$ is the amplitude and $\Psi(f)$ is the phase. The amplitude $A(f)$ is inversely proportional to the effective distance $D_{\rm eff}$, which equals the physical luminosity distance $D$ for optimal source sky location and orientation but otherwise exceeds $D$. The phase $\Psi(f)$ depends on the intrinsic parameters common to all detectors: detector-frame chirp mass $\mathcal{M}_c$, symmetric mass ratio $\eta=M_1\,M_2/(M_1 + M_2)^2$, aligned spin components $s_{1z}$ and $s_{2z}$. At each detector, $h(f)$ further depends on three extrinsic parameters: the effective distance $D_{\rm eff}$, a phase constant $\phi_c$, and an arrival time $t_c$. These parameters are not independent between detectors, but for loud events it is an excellent approximation to fit those separately~\cite{Roulet:2018jbe,Dai:2018dca}.  

For NS mergers, we uses the augmented model \texttt{IMRPhenomD\_NRTidal}~\cite{Dietrich:2017aum, Dietrich:2018uni}. This model includes tidally induced phasing. The reason to use realistic waveform models for proof of concept is to show that diffraction signatures cannot be fully mimicked by a change in intrinsic and extrinsic parameters.

\subsection{Mock Forward-Viterbi tests}
\label{sec:simres}

\begin{figure}[t]
  \begin{center}
    \includegraphics[scale=0.6]{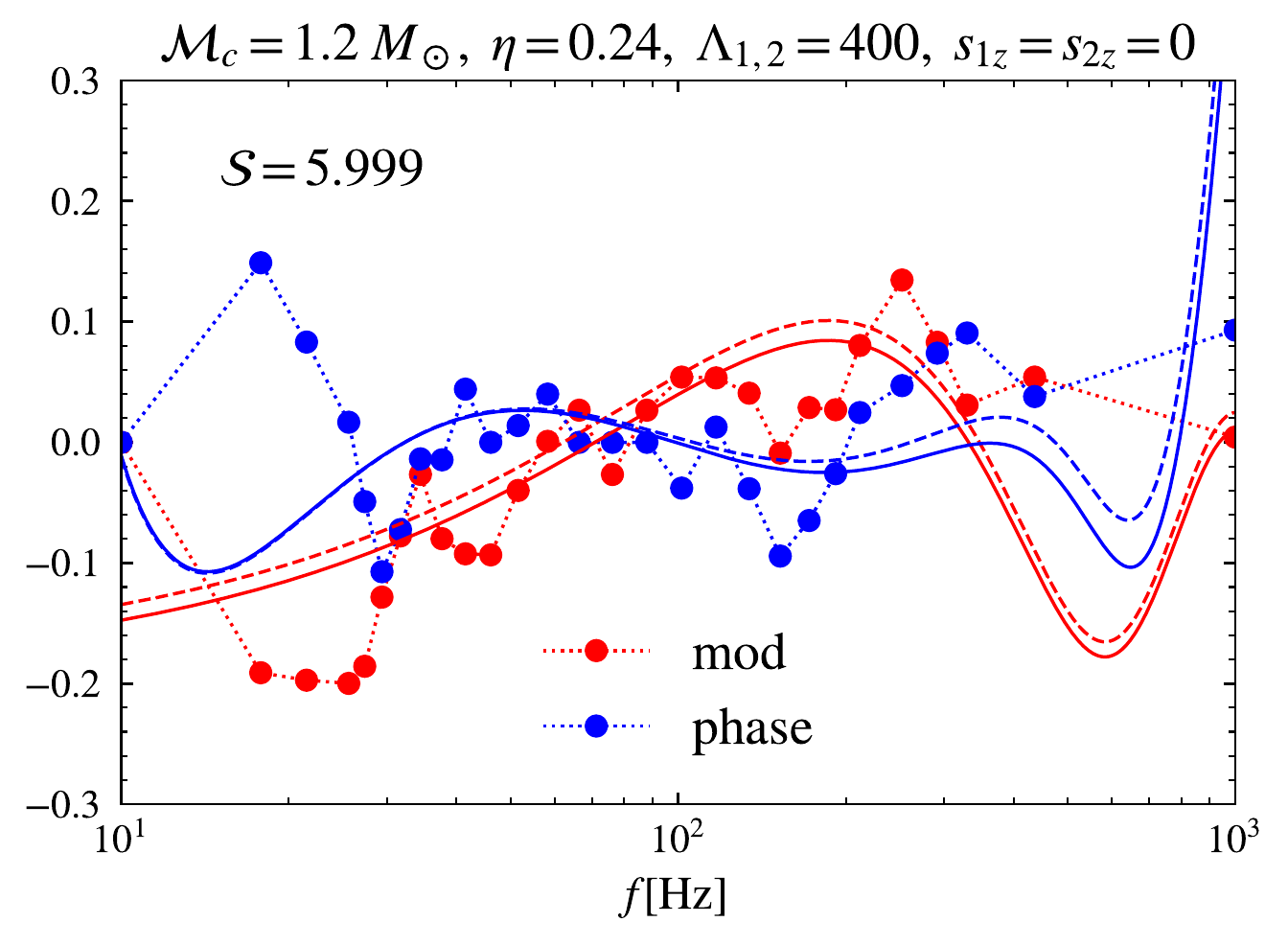}\\
    \hspace{-0.3cm}\includegraphics[scale=0.6]{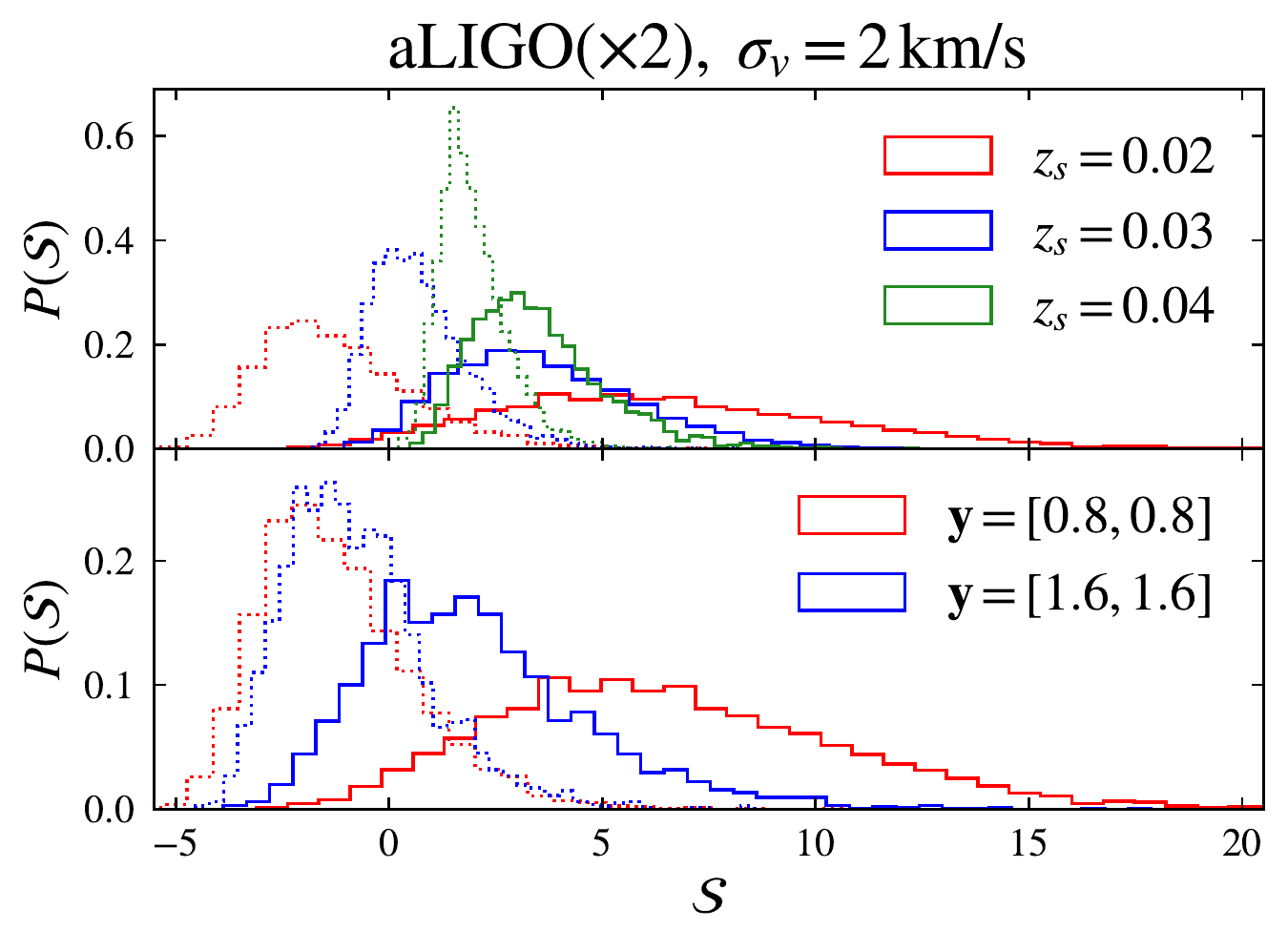}
    \caption{\label{fig:forward_BNS}  A double NS merger detected by two aLIGO detectors at design sensitivity. {\it Upper}: amplitude (fractional) and phase perturbations around the best-fit unlensed template for one noise realization. Refer to the text for the parameters we use. Dots are frequency binned reconstruction from the Viterbi algorithm. Curves (solid and dashed for the two aLIGO detectors respectively) are the theoretical modulation signals computed from $h_L(f)/h_{\rm BF}(f)$. {\it Lower}: Distribution of the score $\mathcal{S}$ with (solid) and without (dotted) diffraction. We show the effect of increasing the source distance (upper panel; assume $z_l = z_s/2$, and for both detectors $D_{\rm eff} = D = 87,\,132,\,177\,$Mpc for $z_s = 0.02,\,0.03,\,0.04$, respectively) and increasing the impact parameter (lower panel; fix $z_s = 0.02$). }
  \end{center}
\end{figure}

Let us consider the two aLIGO detectors detecting a non-spinning double NS merger at their design sensitivities. For demonstration, we choose a chirp mass $\mathcal{M}_c = 1.2\,M_\odot$, a symmetric mass ratio $\eta = 0.24$, and tidal deformability parameters $\Lambda_1 = \Lambda_2 = 400$. We assume the source located at $z_s = 0.02$, $D_{\rm eff} = 87 \,$Mpc for both detectors, and a lens as in Case (1) of \reffig{Fwrel_example} with $\sigma_v = 2\,{\rm km/s}$ and $z_l = 0.01$. 

To apply the Forward-Viterbi test, we divide the frequency range $[10,\, 1024]\,$Hz into 27 frequency bins with nearly equal contributions to the squared matched filtering SNR. Following \refeq{uvgrids}, we limit the fractional distortion in $h(f)$ to $u_{\rm max} = v_{\rm max} = 0.2$ and $u_{\rm min} = v_{\rm min} = -0.2$, and set the number of grid points to be $n_u = n_v = 32$. Furthermore, we set $\Delta k_{\rm max} = \Delta l_{\rm max} = 4$, restricting any ``jump'' between adjacent frequency bins to be within 4 grid points.

The top plot of \reffig{forward_BNS} shows the reconstruction of the diffraction signature for one random noise realization. The Forward algorithm yields a score $\mathcal{S} = 5.999$, which is significantly higher than the typical score one would obtain in the absence of diffraction distortion. The best-fit modulation obtained through the Viterbi algorithm is noisy but on average tracks the underlying signal. As expected, the reconstruction is the most accurate within the frequency range of the highest sensitivity $[30,\,200]\,$Hz .

 The method does not recover $F_{\rm rel}(f)$, but only the part that is not degenerate with the physical parameters. Although it would be difficult to undo this degeneracy, one can still infer from the partial reconstruction of $F_{\rm rel}(f)$ the modulation frequency scale (i.e. the ``oscillation period'' in frequency space), whose inverse connects to the Schwarzschild time scale of the lens.

The measured value for $\mathcal{S}$ should be compared to the distribution of $\mathcal{S}$ under a given hypothesis to be tested. The distribution can be numerically derived by injecting a signal waveform into the noise. The lower plot of \reffig{forward_BNS} compares the distribution of $\mathcal{S}$ between two cases: (1) the diffraction distorted waveform $h_L(f)$ hidden in the noise; (2) the best-fit undistorted waveform $h_{\rm BF}(f)$ hidden in the noise. The less the two distributions overlap, the more detectable the diffraction signature is. 

Two separated distributions may still have non-negligible overlap in the tails. The detection significance is subject to stochasticity due to random noise. For the example cases we show in the lower plot of \reffig{forward_BNS}, it is not always possible to claim a detection even for small source distances, although at a large fraction of the times one would be able to rule out the null hypothesis. The plot demonstrates how detectability degrades as the source distance increases, and as the impact parameter grows. The results suggest that, for an impact parameter on the order of the angular Einstein scale $\theta_E$, a pair of aLIGO detectors at the design sensitivity are sensitive to diffraction signals imprinted in binary NS merger waveforms out to $D_{\rm eff} \simeq 100\,$Mpc.

\begin{figure}[t]
  \begin{center}
    \includegraphics[scale=0.6]{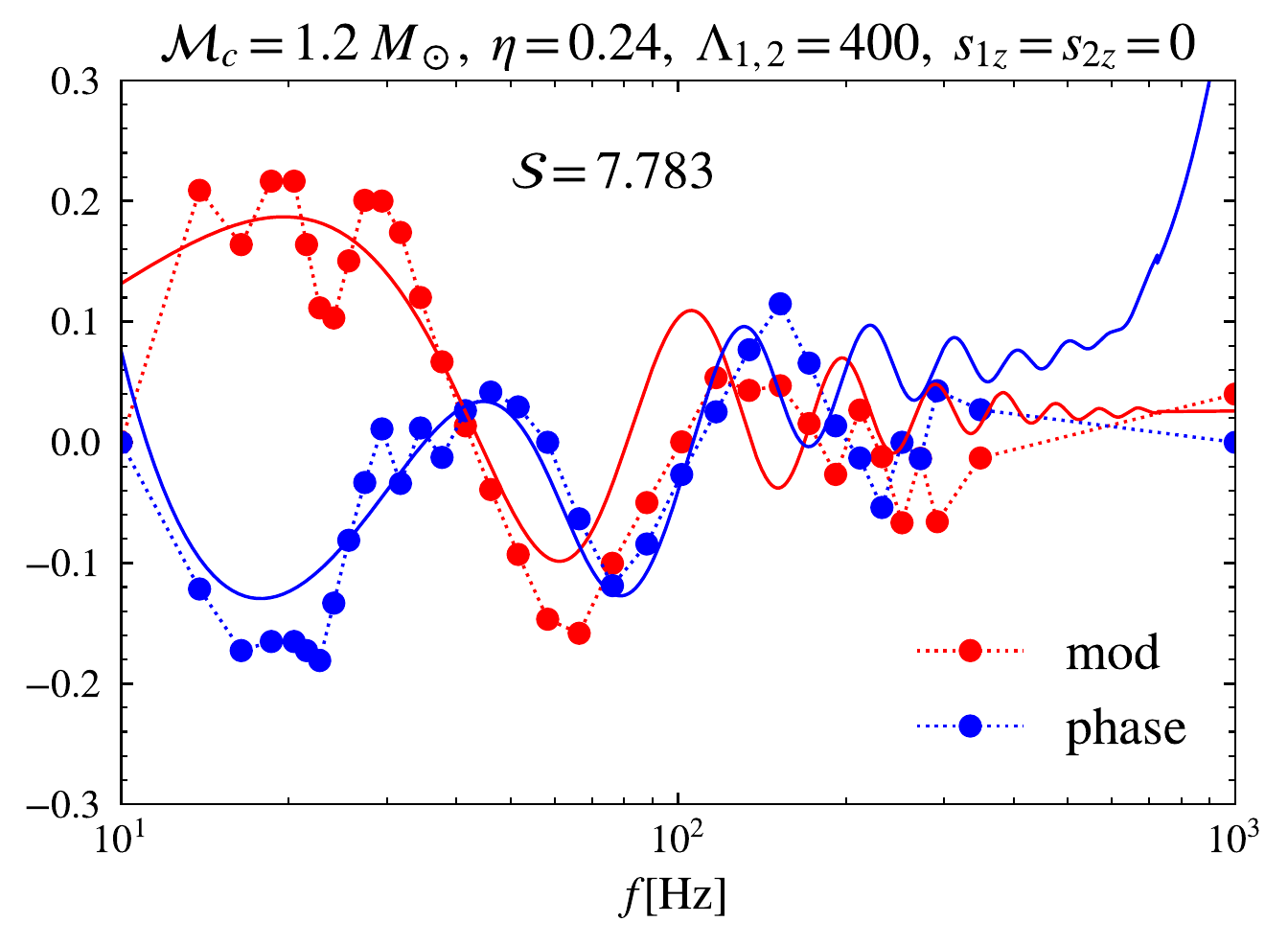}\\
    \hspace{-0.3cm}\includegraphics[scale=0.6]{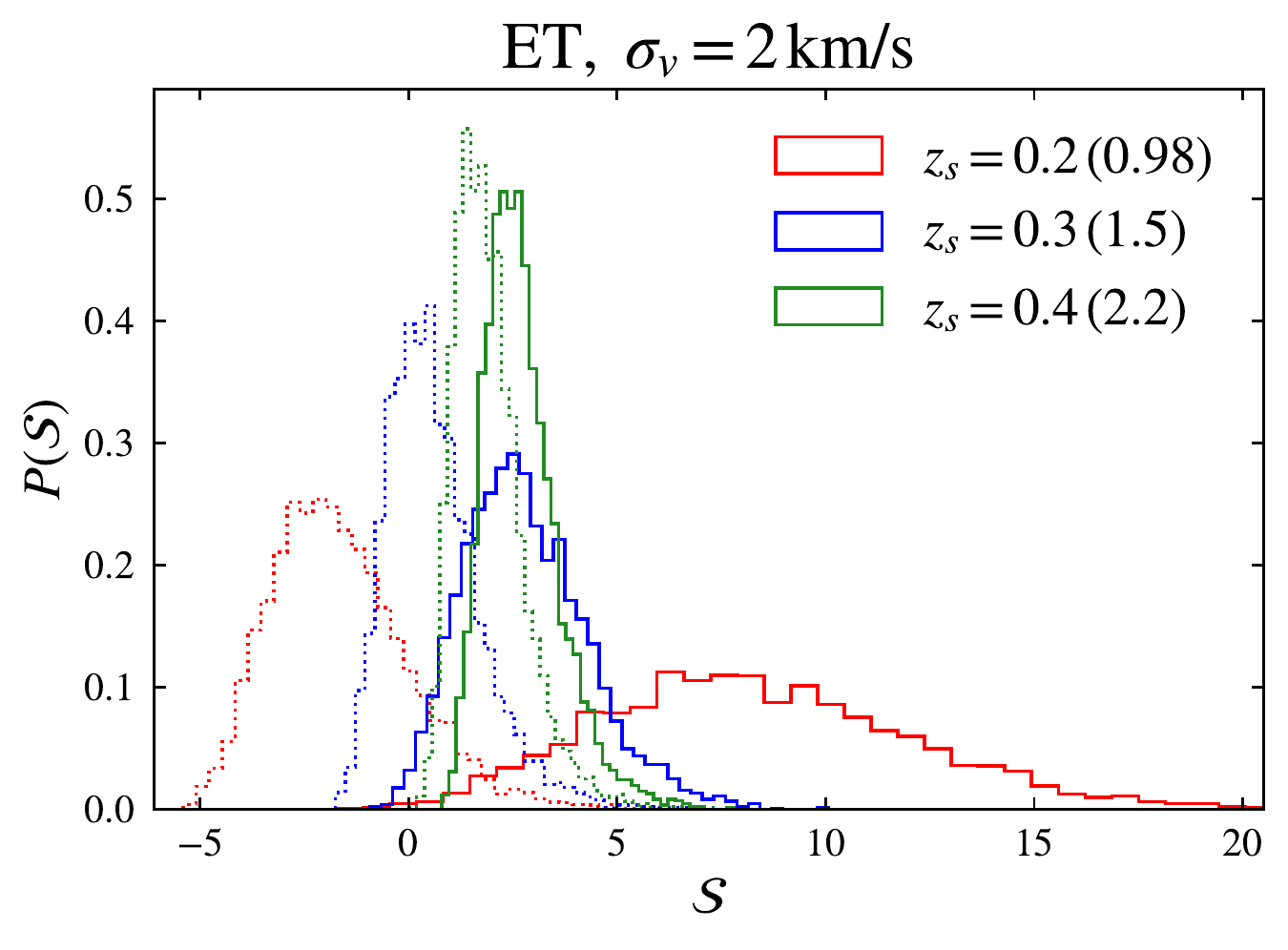}
    \caption{\label{fig:forward_BNS_ET} Same as \reffig{forward_BNS} but for the proposed ET (single detector) and for larger source distances. In the upper plot we assume $z_s = 2\,z_l  =0.2$ and $D_{\rm eff} = 0.98\,$Gpc. In the lower plot, we indicate the effective distance $D_{\rm eff}$ in Gpc between parentheses following the legend label for the source redshift.}
  \end{center}
\end{figure}

\begin{figure}[t]
  \begin{center}
    \includegraphics[scale=0.6]{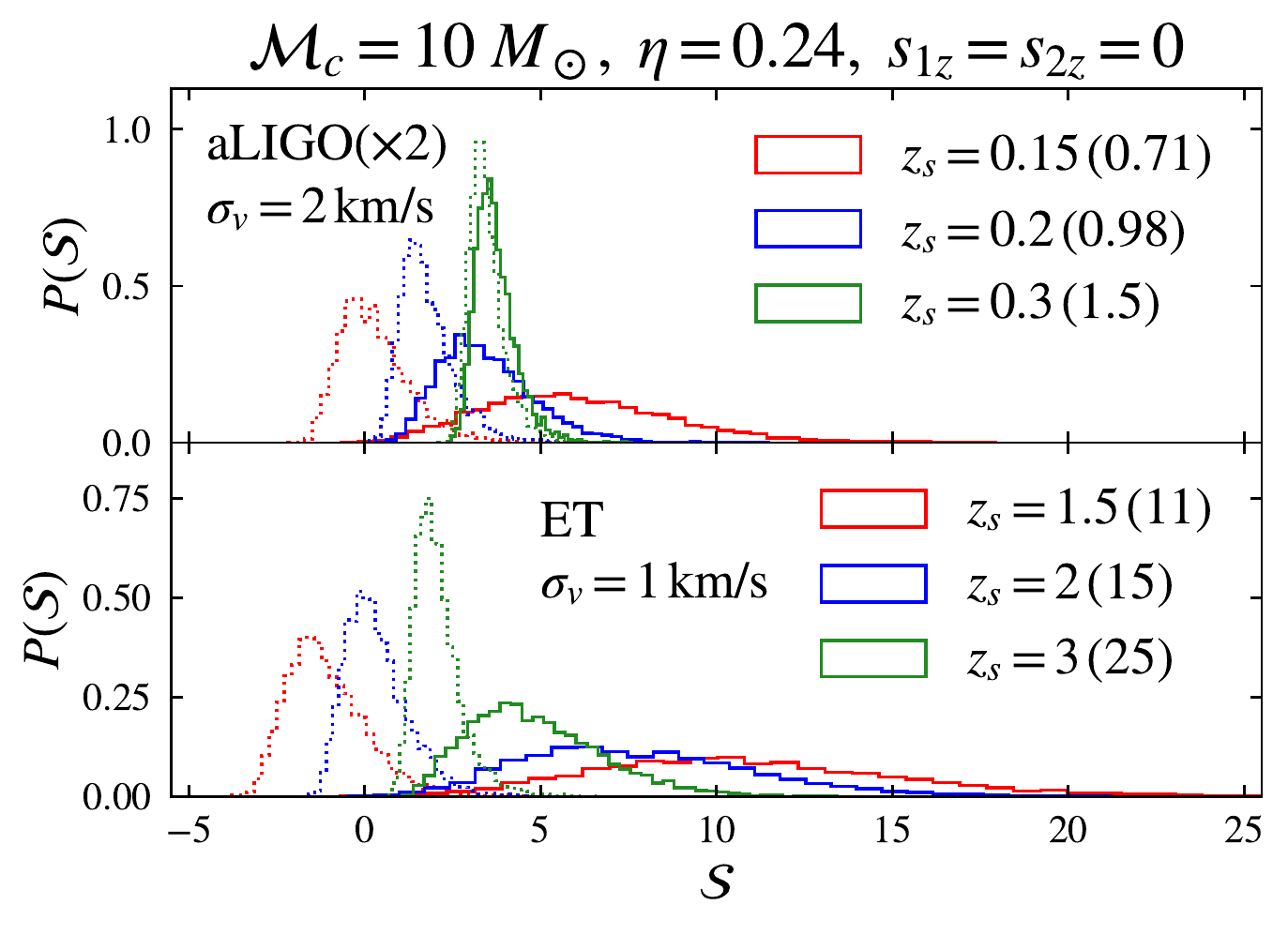}\\
    \includegraphics[scale=0.6]{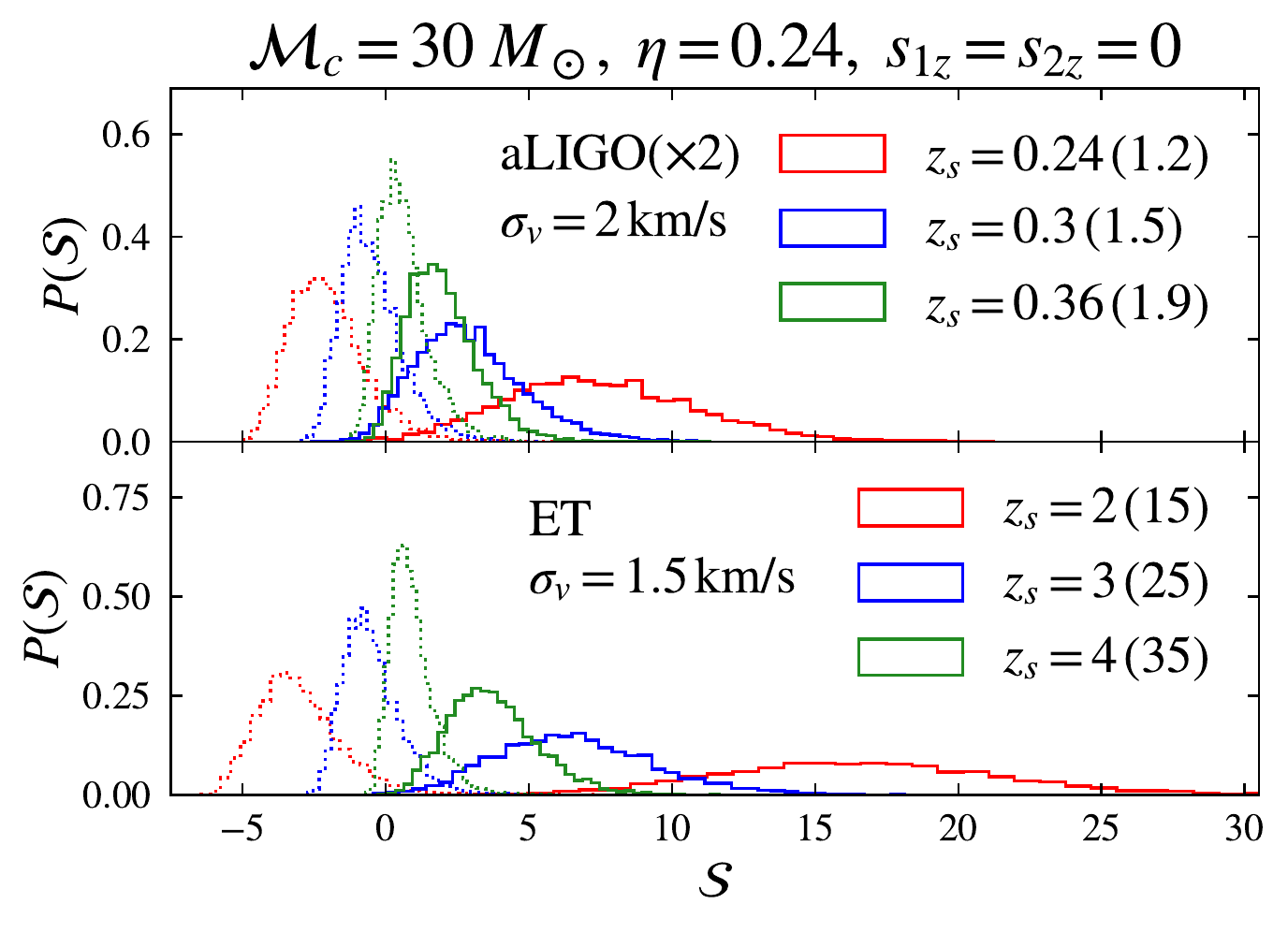}
    \caption{\label{fig:forward_BBH_aLIGO_ET} Distribution of the score $\mathcal{S}$ with (solid) and without (dotted) diffractive lensing for binary BH coalescences. We simulate for comparable component BH masses $\eta = 0.24$ with (source-frame) a chirp mass $\mathcal{M}_c = 10\,M_\odot$ (top plot) and $\mathcal{M}_c = 30\,M_\odot$ (bottom plot), and assume zero spins. In each plot, we consider the case of two aLIGO detectors at the design sensitivity (upper panel) and the case of one third-generation detector as proposed for the ET (lower panel). We fix $z_l = z_s / 2$. We indicate the effective distance $D_{\rm eff}$ in Gpc between parentheses following the legend label for the source redshift.}
  \end{center}
\end{figure}

The range of GW detection will be greatly extended by third-generation detectors. For the same lens we have assumed in the above, the proposed Einstein Telescope (ET) will enable a search for diffraction signature in typical double NS merger events out to $D_{\rm eff} \sim 1\,$Gpc, as shown in \reffig{forward_BNS_ET}.

Next, we apply the same analysis to binary BH mergers. While spanning a smaller frequency range in the ground-based band, they are louder sources than neutron stars. Detectable to larger distances, those can be more efficient probes of lenses along the line of sight. We again use $\sim 30$ frequency bins, but we have adjusted the frequency binning according to how the distribution of the SNR in the frequency domain varies.

The top plot in \reffig{forward_BBH_aLIGO_ET} considers binary BH systems with (source-frame) chirp masses $\mathcal{M}_c = 10\,M_\odot$, whose progenitors may be the observed high mass X-ray binaries. With two aLIGO detectors at the design sensitivity and for the same fiducial lens we have been assuming, diffraction modulations are detectable out to $z_s \sim 0.15$--$0.2$, corresponding to $D_{\rm eff} \sim 0.7$--$1\,$Gpc. This distance could increase by an order of magnitude to $D_{\rm eff} \sim 10$--$20\,$Gpc with just one third-generation detector, potentially reaching binary BH mergers from $z_s \sim 1$--$2$.

The bottom plot in \reffig{forward_BBH_aLIGO_ET} considers more massive binary BH systems with $\mathcal{M}_c = 30\,M_\odot$. Those intriguing systems were first uncovered in GW detections. Due to low cut-off frequencies, those are limited by the frequency span that can adequately sampled, which will further exacerbate for highly redshifted systems. Detectable diffraction-induced modulation thus must fall within the right frequency range that has a high SNR. Despite that, the strong GW power from those systems still make them suitable sources for probing intervening lenses out to very large distances. Two detectors at the fully upgraded aLIGO will reach $z_s \sim 0.25$ or $D_{\rm eff} \sim 1\,$Gpc. Tremendous improvement can be expected for third-generation detectors. The ET will enable to utilize sources out to $z_s \sim 2$--$4$ or $D_{\rm eff}\sim 15$--$35$\,Gpc.

\subsection{Comparison to matched filtering}
\label{sec:comparemf}

\begin{figure*}[t]
  \begin{center}
    \includegraphics[scale=0.6]{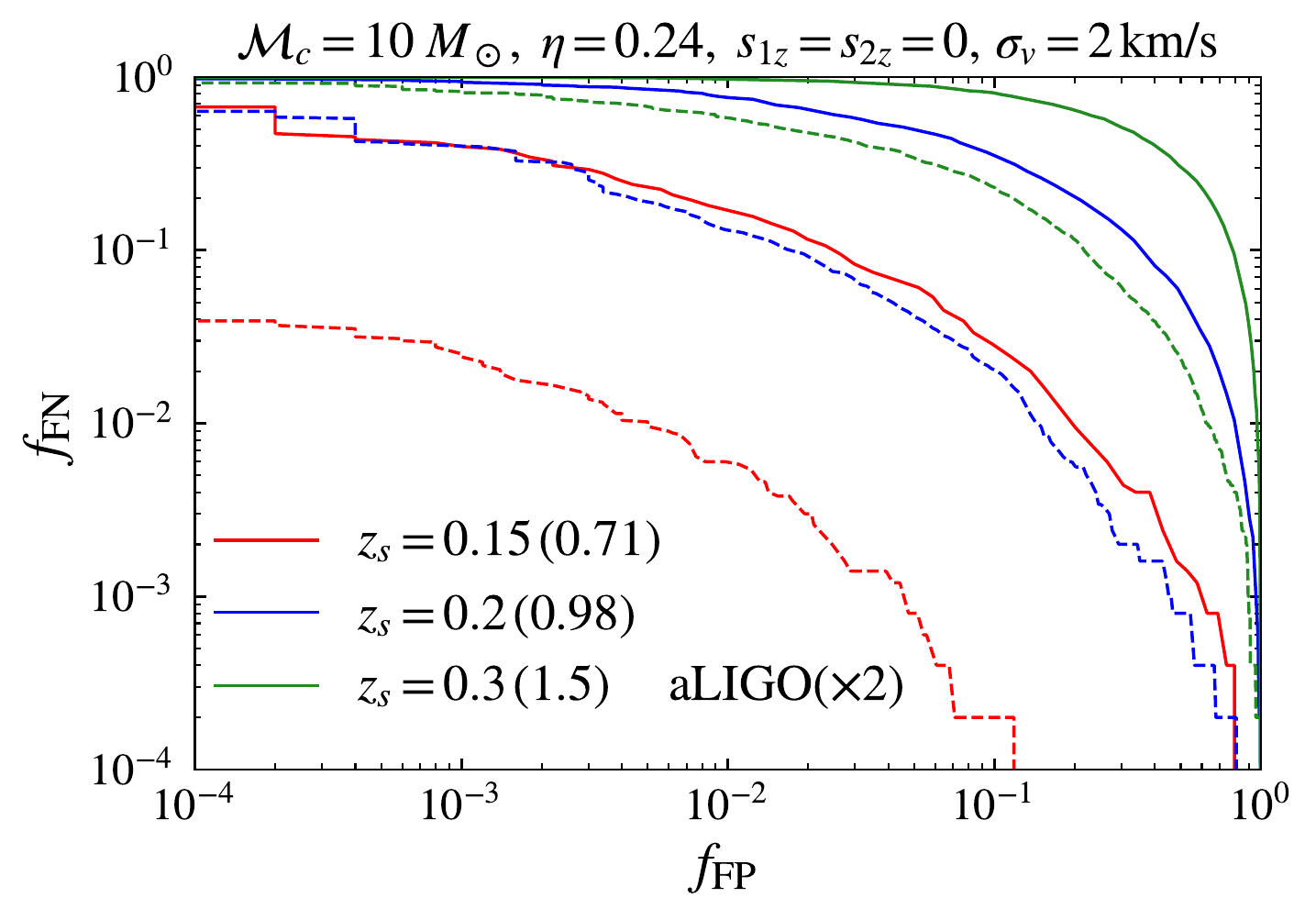}
    \includegraphics[scale=0.6]{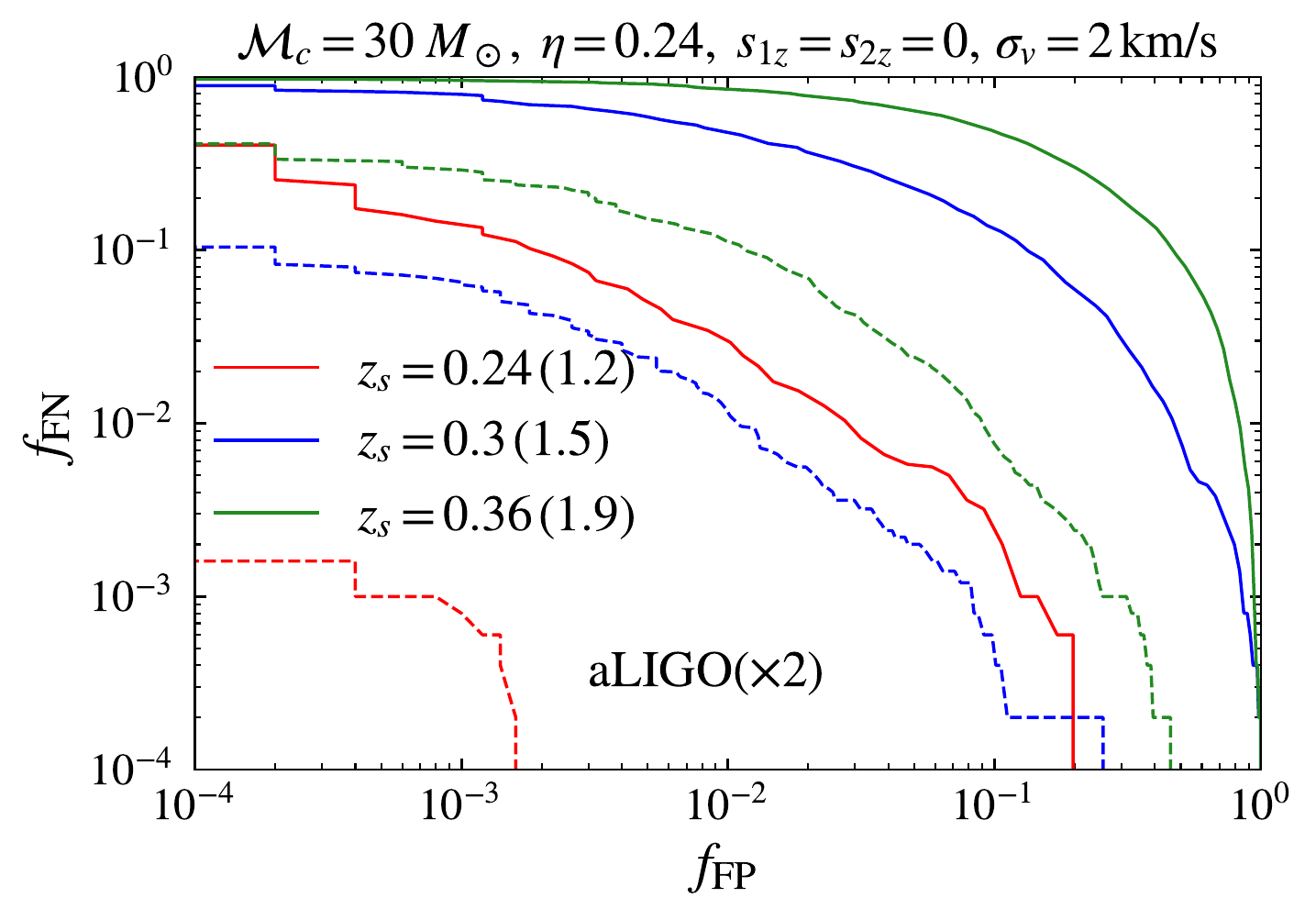}
    \includegraphics[scale=0.6]{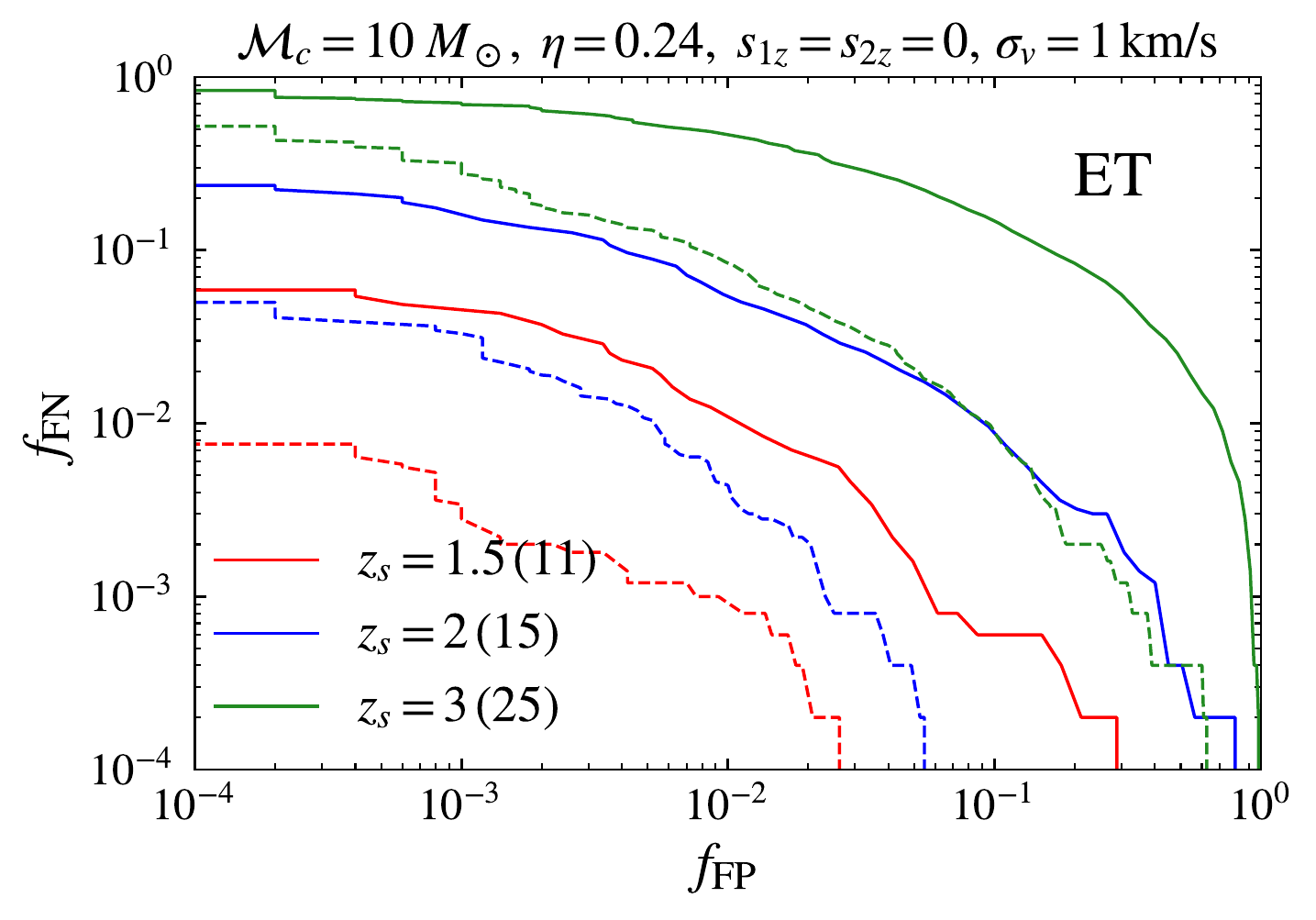}
    \includegraphics[scale=0.6]{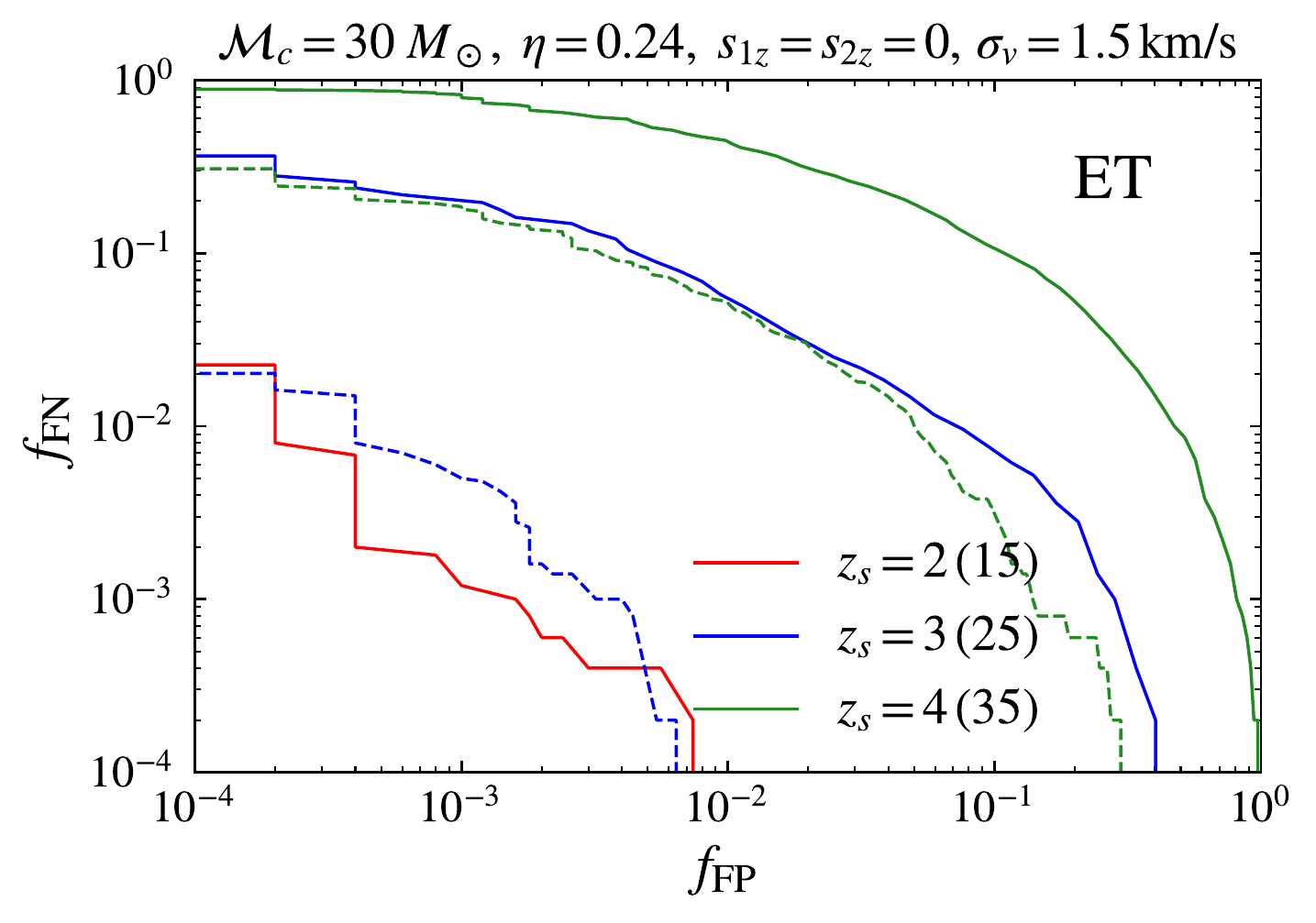}
    \caption{\label{fig:FP_vs_FN} Relation between the false positive probability $f_{\rm FP}$ and the false negative probability $f_{\rm FN}$ for a single GW event. We consider binary BH mergers, with the same parameters as used in \reffig{forward_BBH_aLIGO_ET}. We compare dynamic programming (using the score $\mathcal{S}$ of \refeq{mcalSdscr}; solid curves) to matched filtering (using the score $\mathcal{S}_{\rm mf}$ of \refeq{calSmf}; dashed curves). The legends indicate the source redshift $z_s$ followed by the effective distance $D_{\rm eff}\,$[Gpc] given between parentheses. We assume two aLIGO detectors at the design sensitivity for the top plots, and one third-generation detector for the bottom plots. In the bottom right plot, the curve for matched filtering for the case $z_s = 4$ is not shown because $f_{\rm FP}$ and $f_{\rm FN}$ are so small that the sample size of our mocks is insufficient.}
  \end{center}
\end{figure*}

We now compare dynamic programming to matched filtering. For the latter, we define a score
\ba
\label{eq:calSmf}
\mathcal{S}_{\rm mf} := \frac12\,\sum^{N_d}_{a=1}\,\left[ \frac{\left|\left(s_a | \tilde{h}_{\rm BF, a} \right)\right|^2}{\VEV{\tilde{h}_{\rm BF, a}|\tilde{h}_{\rm BF, a}}} - \frac{\left|\left(s_a | h_{\rm BF, a} \right)\right|^2}{\VEV{h_{\rm BF, a}|h_{\rm BF, a}}} \right],
\ea
which quantifies the improvement in the log-likelihood function after diffractive distortion is allowed into the waveform model. Here $h_{\rm BF, a}(f)$ is the best-fit {\it unlensed} waveform, and $\tilde{h}_{\rm BF, a}(f)$ is the best-fit diffraction-distorted waveform in the form of an unknown unlensed waveform multiplied by $F_{\rm rel}(f)$. We pretend that {\it the true $F_{\rm rel}(f)$ is exactly known}. Compared to the idealistic analysis of \refsec{mf} (c.f. \refeq{lnp}), \refeq{calSmf} accounts for observational degeneracy between diffraction and the source parameters, and allows stochasticity from the detector noise.

Similar to testing out dynamic programming, we can derive the distribution for the score, both in the presence of diffraction and under the null hypothesis. For a {\it single} GW event, and at a given threshold value $\mathcal{S}_c$ for claiming a detection, the false positive probability is given by the cumulative distribution $f_{\rm FP}:= P_0(\mathcal{S} > \mathcal{S}_c)$ computed under the null hypothesis, and the false negative probability is given by $f_{\rm FN} := P(\mathcal{S} < \mathcal{S}_c)$ computed in the presence of diffraction. One way to characterize the effectiveness of a given score is to map out a relation between $f_{\rm FP}$ and $f_{\rm FN}$ by continuously varying $\mathcal{S}_c$.

Taking binary BH mergers as an example, we show curves for $f_{\rm FP}$ versus $f_{\rm FN}$ in \reffig{FP_vs_FN} using the distributions presented in \reffig{forward_BBH_aLIGO_ET}. In particular, we make a comparison between our implementation of dynamic programming and matched filtering. Compared to matched filtering, dynamic programming is much more practical when $F_{\rm rel}(f)$ is not known. However, the advantage of being agnostic comes at the expense of large reduction in sensitivity relative to matched filtering. Consequently, the horizon distances we have found for dynamic programming are necessarily smaller than the na\"{i}ve estimates of \reffig{snr_Deff}. For the same parameters we have chosen for the Forward-Viterbi filter, the false positive rate typically worsens by one or two orders of magnitude relative to matched filtering at a fixed false negative rate $f_{\rm FN} \sim 10$\%. 

The optical depth to diffractive lensing of distant sources is likely to be small (see discussion in \refsec{concl}). Only after many GW events are analyzed, one of them may be found to exhibit non-trivial waveform distortions. If diffractive lensing occurs once among every thousand events, and if we take the simplifying assumption that all events are similar, we will have to achieve an expected single-event false positive probability with our detection method that is substantially less than $10^{-3}$. Since the lensing optical depth grows quickly from $z_s \sim 0.2$ to $z_s \sim 2$--$3$, this penalty may be an order of magnitude more severe for second generation detectors than for third generation detectors.

For the above reason, it is of great importance to optimize the Forward-Viterbi filter for a smaller single-event false positive probability. The knowledge of the $F_{\rm rel}(f)$'s generic behavior is crucial for designing the best frequency binning, the best discretization scheme for $F_{\rm rel}(f)$, and the best prior function $\calP[{u_j, v_j}]$, all of which would help mitigating the look-elsewhere penalty. A more dedicated study of this optimization problem goes beyond the scope of this work. We defer such a study to future work. 

We have only quoted results for a pseudo-Jaffe lens, with a specific choice for the impact parameter on the order of the Einstein radius. Detectability of the diffraction signature may substantially vary depending on the lens profile, the impact parameter, and the influence of external convergence and shear. In the case of a low mass lens embedded in a massive lens as substructure, external convergence and shear can amplify the diffraction-induced modulations. Further work is in need for a thorough exploration of the parameter space. 

\section{On precessing spins and eccentricity}
\label{sec:issue}

The waveform models we have used for demonstration are highly realistic but are not fully general. Waveforms describing compact binary coalescence can exhibit imprints from spin-orbit precession due to misaligned spins and from orbital eccentricity.

The effects of binary masses, aligned spins, and tidal deformabilities are distinguishable from diffraction induced modulations. This is because those do not cause oscillations in the amplitude and in the unwrapped phase of the frequency-domain waveform. However, this is not the case for binaries with precessing spins. The GW amplitude oscillates as the orbital plane wobbles around the direction of the total angular momentum vector on a timescale that is $\mathcal{O}[(v/c)^{-2}]$ longer than the orbital timescale. For systems suitable for ground-based detection, the orbital plane wobbles for about $\mathcal{O}(10)$ cycles through the band, a number largely insensitive to spin magnitudes~\cite{PhysRevD.49.6274}. Precessing spins also cause small oscillations in the unwrapped waveform phase~\cite{Klein:2013qda, chatziioannou2017analytic}. 

Misaligned spins are certainly possible for physical binary mergers~\cite{Harry:2016ijz}. Their effects on the waveform, however, may not be severely degenerate with diffraction. While the latter creates modulation cycles linearly spaced with the frequency, precession modulations are more densely packed toward low frequencies. We have seen that the first diffraction peak at low frequencies is the foremost target for detection, while for spin-orbit precession we would expect many modulation cycles in the same frequency range. Moreover, spin-orbit precession tends to induce an amplitude modulation that is significantly greater in size than the phase modulation~\cite{Klein:2013qda}. For diffraction the two would have comparable sizes. The oscillatory effects should also be fit simultaneously with the non-oscillatory phasing corrections induced by misaligned spins. 

The issue may also be relevant if non-zero orbital eccentricity is allowed. In this case, oscillation occurs on the timescale of relativistic periastron precession, which is again $\mathcal{O}[(v/c)^{-2}]$ longer than the orbital period. This induces rather rapid modulation cycles in the frequency domain~\cite{Hinder:2017sxy}. Eccentric binaries also distribute their GW power into higher harmonics, a feature not present with diffraction.

Further details need to be worked out for quantifying how waveform modulation from precessing spins and eccentricity may resemble that from diffraction, for which accurate and efficient frequency-domain waveform templates are crucial. When applying dynamic programming, one should first find the best-fit unlensed waveform with the extended waveform model, and then seek additional perturbations around the best-fit solution using dynamic programming. In order to mitigate possible degeneracy with other sources of waveform modulation, we may design special prior in favor of diffraction-like distortions.

\section{Discussion}
\label{sec:concl}

This work has focused on the feasibility of probing intervening low mass DM clumps through their diffractive lensing effects imprinted in astrophysical GWs detectable at ground-based detectors. The frequency coverage of aLIGO/Virgo and their forthcoming companion observatories translates into a lens mass scale $\sim 10^2$--$10^3\,M_\odot$ enclosed within a radius of order the impact parameter. The sensitivity to low mass halos will be useful in differentiating warm and cold dark matter scenarios~\cite{2016MNRAS.460..363L}. 

We have developed a dynamic-programming-based algorithm to search for amplitude and phase modulations imprinted in the waveform due to diffractive lensing. Unlike matched filtering, the algorithm does not require a template bank for lensed waveforms. It is a practical and computationally cheap method which can be straightforwardly incorporated into the current framework of GW data analysis. While being sub-optimal compared to matched filtering (if the exact lensed waveform model is known), the method allows to properly quantify the look-elsewhere penalty of trying out a large number of possible waveform distortions.

We have demonstrated the general feasibility of our method using mock detections of injected GWs. We have verified that the diffraction signature is not completely degenerate with many of the binary parameters, including the masses, the aligned spins, tidal phasing, the arrival time, the phase constant, and the overall amplitude normalization. Future work should shed light on whether diffraction modulation can be degenerate with the effects of spin-orbit precession and orbital eccentricity.

We assessed detectability assuming a fiducial pseudo-Jaffe lens with an impact parameter on the order of the Einstein radius, and found that the range of detectability can be interesting for binary BH mergers. Two fully upgraded aLIGO detectors can jointly probe diffraction imprints using binary BH mergers out to $\gtrsim 1\,$Gpc or $z_s \sim 0.2$--$0.3$. Third-generation detectors will be much more powerful for this test. Just a single ET-like detector will enable to utilize binary BH sources to probe lenses out to $\gtrsim 10\,$Gpc or $z_s \gtrsim 2$. Such large source distances are much more favorable for the line of sight to intersect any intervening halo.

We note that detectability may vary substantially depending on the lens profile and the impact parameter, which alters the modulation size. Without a specific theoretical prediction for the mass profile of the low mass halos, we have not attempted to thoroughly chart the parameter space. For our fiducial lens model, our estimates correspond to an impact parameter on the order of the Einstein radius.

What might be the probability for diffractive lensing to occur in a CDM universe? A quick estimate might start with the assumption that all DM is locked up in halos of various masses, say $M_L \sim 10^0$--$10^{15}\,M_\odot$, with a mass function such that equal logarithmic intervals in the halo mass contribute the same mass (as is nearly the case for substructure mass function inside a cluster or galactic halo~\cite{mo2010galaxy}). If the observationally relevant halos span one decade in the mass around some characteristic mass scale $M_L$, they account for some fraction $1/\mathcal{N}$ of the total mass in the Universe, where we may take $\mathcal{N} \simeq 15$. For a typical source redshift $z_l$ and proper distance $r$, on average the line of sight intersects one halo at a chance
\ba
\hspace{-0.2cm}
\sim 0.003\,\left(\frac{1+z_l}{2} \right)^3\,\left(\frac{15}{\mathcal{N}}\right)\,\left( \frac{r}{5\,{\rm Gpc}} \right)\,\left( \frac{10^5\,M_\odot}{M_L} \right)\,\left( \frac{b}{1\,{\rm pc}} \right)^2,
\ea
where $b$ is the maximum impact parameter required. A standard NFW halo~\cite{1996ApJ...462..563N} with $M_{200} = 10^5\,M_\odot$ encloses a column of mass $M_{\rm enc}\sim 100\,M_\odot$ within $b = 1\,{\rm pc}$ if it has a concentration $c_{200} = R_{200}/R_s = 30$~\cite{ludlow2016mass}, while the corresponding Einstein radius falls a factor of ten short $r_E = 0.1\,{\rm pc}\,(M_{\rm enc}/100\,M_\odot)^{1/2}\,(d/1\,{\rm Gpc})^{1/2}$, where $d$ is some characteristic angular diameter distance. Note, however, that this probes the region well within the scale radius $b/R_s = 0.04$. If the NFW model underestimates the mass profile slope at small radii for low mass halos~\cite{doi:10.1093/mnras/stu742}, the enclosed mass within the impact parameter may be significantly larger without altering the halo's overall mass scale, leading to a larger Einstein radius and increased strong lensing probability. Halo ellipticity also in general enhances this probability. In any case, the above crude answer suggests that developing third-generation GW detectors are strongly desirable for fully realizing this observational potential.

We further note that theoretically we expect a fraction $\sim 10^{-3}$ among the sources from cosmological distances $z_s \sim 1$--$2$ should be strongly lensed by an intervening galaxy~\cite{Hilbert:2007jd, Takahashi:2011qd, Dai:2016igl, Li:2018prc}. In this case, GWs associated with each macro image propagate through the halo of the lens galaxy and hence has an enhanced probability of intersecting a low mass halo as a substructure. Also, amplified modulation should be expected due to the external shear associated with the macro image (c.f. \reffig{Fwrel_example}). This suggests that GW events subject to galaxy lensing may be promising candidates. Detailed calculations are warranted in the future to assess the observational prospect for any given DM model.

In the regime of our interest, the lensing configuration would not change over the human time scale. If the host galaxy of the GW source can be identified, follow-up imagings may provide a cross check by searching for lensing distortions in the galaxy image~\cite{doi:10.1111/j.1365-2966.2012.20438.x}. At the same level of chance alignment, low mass halos should only cause moderate flux magnification at optical/infrared wavelengths, contributing to the scatter in the apparent luminosity of cosmological standard candles~\cite{Zumalacarregui:2017qqd}. This can provide a multi-wavelength cross check for the lensing effect we seek with GWs.

Although we have considered lensing by DM halos, our technique should be applicable to searching for wave diffraction induced by compact object lensing at large impact parameters, which will extend the work in Ref.~\cite{Jung:2017flg}. This will constrain the abundance of primordial BHs as possible LIGO sources~\cite{Bird:2016dcv, Clesse:2016vqa}.

Finally, it would be interesting to consider GW sources for space-based observatories, extending previous work on the wave effects~\cite{Takahashi:2003ix}. In this case, the space-based frequency band $f \sim 10^{-4}$--$10^{-2}\,$Hz corresponds to very different mass scales $M_E \sim 10^6$--$10^8\,M_\odot$.

\appendix 


\begin{acknowledgments}

The authors are grateful to Hideyuki Tagoshi and Matias Zaldarriaga for useful discussions. LD is supported at the Institute for Advanced Study by NASA through Einstein Postdoctoral Fellowship grant number PF5-160135 awarded by the Chandra X-ray Center, which is operated by the Smithsonian Astrophysical Observatory for NASA under contract NAS8-03060. BZ acknowledges support from the Infosys Membership Fund. This work is also partly supported by the National Key Basic Research and Development Program of China (No. 2018YFA0404501 to SM), by the National Science Foundation of China (Grant No. 11333003, 11390372 and 11761131004 to SM, 11690024 to YL), and by the Strategic Priority Program of the Chinese
Academy of Sciences (Grant No. XDB 23040100 to YL).

\end{acknowledgments}


\bibliographystyle{apsrev4-1-etal}
\bibliography{diffraction}

\end{document}